\documentclass[a4paper]{article}
\usepackage{cite}
\usepackage{hyperref}
\usepackage[pdftex]{graphicx}
\usepackage{url}
\usepackage{xcolor}
\usepackage{multicol}
\usepackage{rotating}
\usepackage{booktabs}
\usepackage[T1]{fontenc}
\usepackage{caption}
\usepackage{subcaption}
\usepackage{multirow}

\graphicspath{{../figures/}{../jpeg/}}
\DeclareGraphicsExtensions{.pdf,.jpeg,.png}

\providecommand{\keywords}[1]
{
	\small	
	\textbf{\textit{Keywords---}} #1
}

\title{Review of Low Voltage Load Forecasting: Methods, Applications, and Recommendations}

\author{Stephen~Haben\\
University of Oxford\\
\and ~Siddharth~Arora\\
University of Oxford\\
\and ~Georgios~Giasemidis\\
Independent Researcher\\
\and ~Marcus~Voss\\
Technische Universit\"at Berlin (DAI-Labor)\\
\and ~Danica~Vukadinovi\'c~Greetham\\
Tessella, Abingdon\\
}

\date{}

\begin{document}
\maketitle
% make the title area

\begin{abstract}
	The increased digitalisation and monitoring of the energy system opens up numerous opportunities to decarbonise the energy system. Applications on low voltage, local networks, such as community energy markets and smart storage will facilitate decarbonisation, but they will require advanced control and management. Reliable forecasting will be a necessary component of many of these systems to anticipate key features and uncertainties. Despite this urgent need, there has not yet been an extensive investigation into the current state-of-the-art of low voltage level forecasts, other than at the smart meter level. This paper aims to provide a comprehensive overview of the landscape, current approaches, core applications, challenges and recommendations. Another aim of this paper is to facilitate the continued improvement and advancement in this area. To this end, the paper also surveys some of the most relevant and promising trends. It establishes an open, community-driven list of the known low voltage level open datasets to encourage further research and development. 
\end{abstract}


\section{Introduction}
%%%
Increased monitoring and communications are opening up opportunities for smart energy networks. The transition to a more localised and distributed energy system helps to support the increased connection of low carbon technologies and provides an environment for new products and services such as peer-to-peer electricity markets, heat-as-a-service, smart storage, and increased renewable generation utilisation. While the uptake of low-carbon technologies (LCTs) is growing and spreading across the globe, predicting their growth in a specific location is challenging. Furthermore, smart technologies to manage LCT's impact on the grid is still in their relative infancy. Low voltage (LV) network modelling will improve network planning and facilitate better management of potential LCT connection hotspots. It could also allow local authorities to monitor their progress toward a more sustainable future. Our ability to facilitate and optimise these opportunities would necessitate access to accurate forecasts of the demand at the low voltage level, given that future estimates of load helps anticipate the core features of LV network models. 

Unfortunately, there is a dearth of literature on forecasting at the low voltage level. Smart meter roll-outs worldwide have increased the investigation into household-level demand \cite{Wang2018ros}, however, very few papers focus on the secondary or even primary substation level of the distribution network. In a handful of existing studies, forecasts on low voltage networks are typically based on aggregations of multiple smart meters. While this is encouraging, it has been shown that this approach does not provide a perfect representation of LV load as it overlooks many important features and nuances of a real LV network \cite{Haben2019stl}.

Although literature and methods are abundant for forecasting at the higher voltage and system levels \cite{Tao2020efa}, due to the increased volatility at the low voltage level, other challenges, not present at the system level, emerge.
Given these reasons, we need more advanced methods to accommodate a more complex range of patterns in energy time series at the LV level, whereby the underlying uncertainty is also communicated to the end-user in the form of probabilistic forecasts. There is a requirement for further research in this area that spans more advanced and complex methodologies. 

This paper serves to review the current research of load forecasting at the low voltage level, to identify the gaps and opportunities, highlight the challenges, and provide recommendations and best practices.

\begin{figure*}
	\includegraphics[,width=\textwidth]{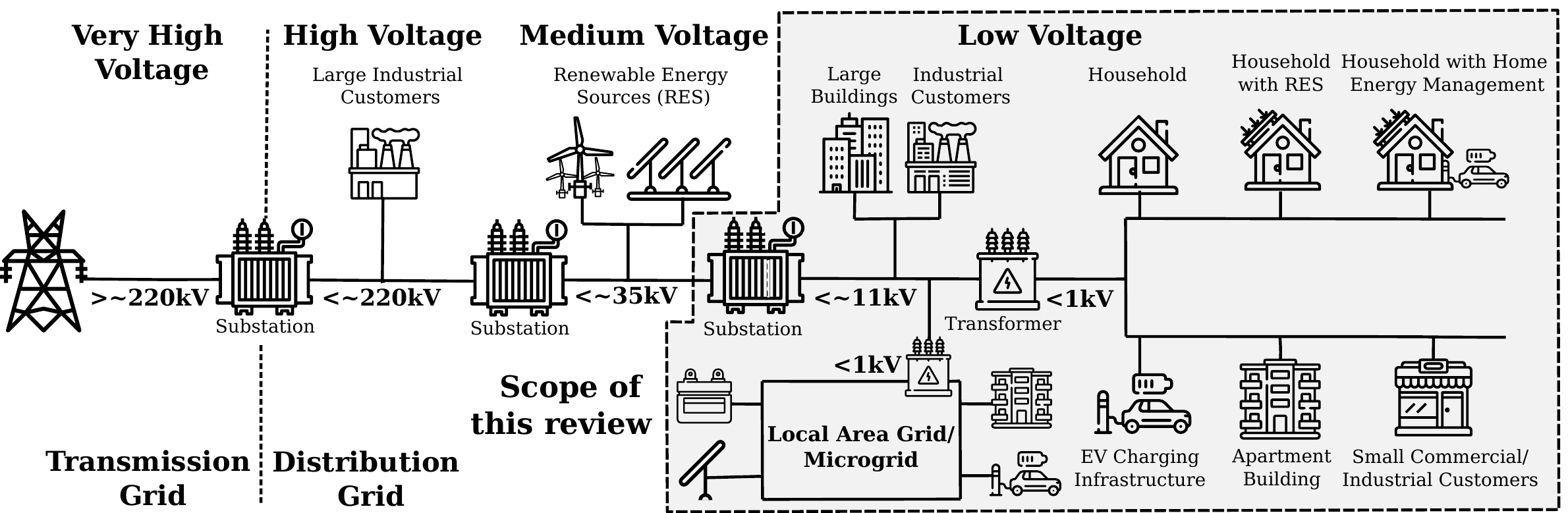}\par 
	\caption{Overview of typical high, medium and low voltage grid layout. This shaded area shows the scope of this review. }
	\label{fig:lvgrids}
\end{figure*}

Low voltage can be an ambiguous term, and hence before proceeding, it is worth defining the scope of this review. Low voltage is a relative term and is defined differently in different countries. Figure~\ref{fig:lvgrids} gives a simplified overview of the layout of electrical grids, as they are typical, for instance, in the UK and Europe. The voltages are stepped down via several substations from the very high transmission level voltages of, for example, 220kV or 300kV in Germany and up to 400kV in the UK, down to the 400V or 230V at the end customer level. The high voltage level of distribution grids has typically voltages of, for instance, 110kV (Germany) or 132 kV (UK), followed by a medium voltage level that ranges from 1 kV up to 35 kV. Given the large number of distribution systems, the layout at this medium voltage level varies across regions, even within countries. This level incorporates the 11kV and 33kV levels common in the UK, typical medium voltages levels of 10-20 kV in Germany, and the corresponding parts in US distribution systems rated at, for instance, 34.5kV, 13.2kV, and 4.16kV. 

For the scope of this review, we also include the lowest level of the medium voltage range, typically at 11kV in the UK and 10kV in Europe, down to the end-customer level. This ensures that the review remains focused on the challenging area of the "last mile'' of the distribution network with its numerous applications, heterogeneous end-customers, and relatively high volatility. The scope further retains forecasting within local area grids and microgrids that are operated at low voltages, as well as the special case of the household level.\footnote{In some cases it is not obvious whether a paper is LV by our definition and in this case, the paper is still reviewed and included as it is still at distribution and not transmission level.}

\subsection{Motivation}
\label{sec_motivation_relatedrevs}

Until recently, the focus of most research on short term load forecasting has been on the system level with the aim of balancing supply and demand \cite{Tao2020efa}. As discussed in \cite{diamantoulakis2015bda}, this was based on large aggregated load data where the individual variations have been averaged out. With more distributed generation, the focus is moving towards more consumer-centric models. The challenges of the future low carbon network are likely to be increasingly concentrated at the low voltage level and decentralisation in general.  

Moving forward, this will mean more locally-focused analytics, i.e. the so-called \textit{smart grid applications}, where capabilities such as smart storage control, demand-side response and peer-to-peer markets are expected to play a major role. A smart grid essentially uses advanced metering with two-way
communications to monitor, detect and respond to energy usage. Forecasting will be an essential component of smart grids since they will allow networks to anticipate, and hence prepare for significant changes in demand. Hence, although many of the applications defined in Section~\ref{sec:LVLF-applications} consider components of a smart grid, the smart grid is not synonymous with the low voltage network, the focus of this review.

There are a growing number of papers on load forecasting \cite{Tao2016pel, Tao2020efa}, however, contributions are mostly restricted to either very large aggregations, typically system-level loads, or at the individual household level, with a particular focus on smart meter data. The LV distribution level which lies in the middle of the network hierarchy has unfortunately received very little attention, hence this review. 

\begin{figure*}
	\begin{multicols}{2}
		\includegraphics[clip, trim=0.5cm 9cm 0.5cm 9.25cm, width=0.48\textwidth]{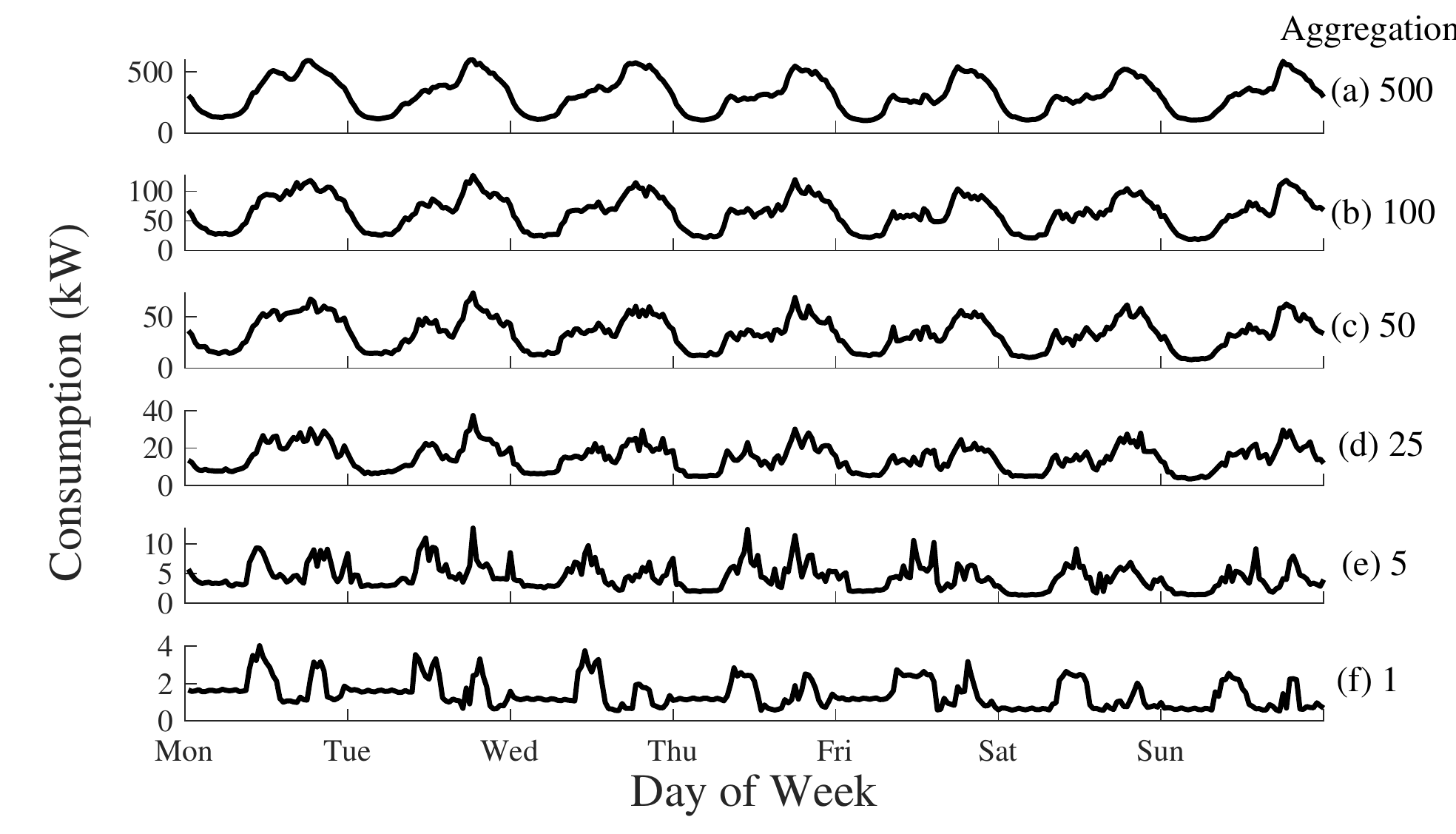}\par 
		\includegraphics[clip, trim=0.5cm 9cm 0.5cm 9.25cm, width=0.48\textwidth]{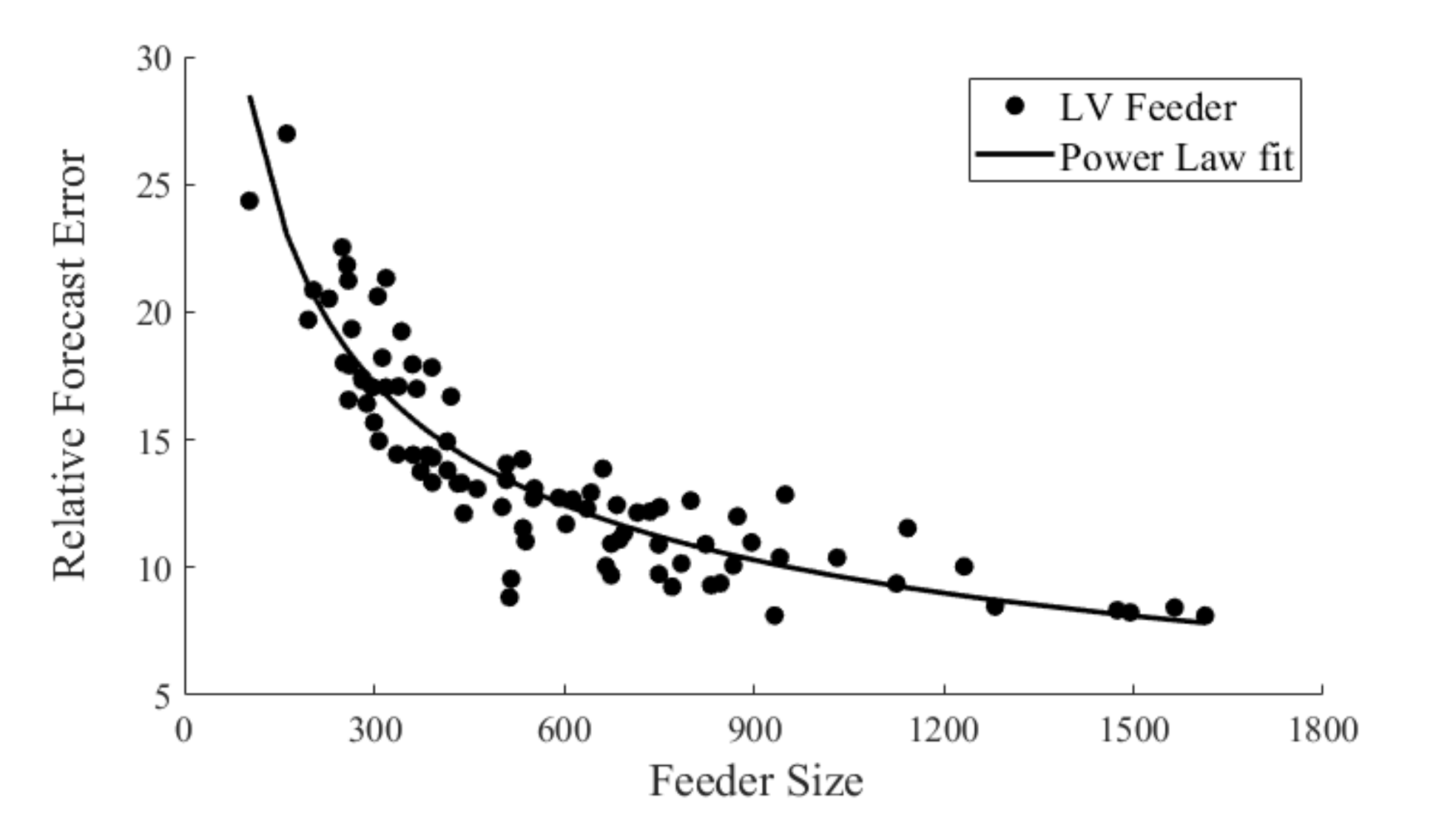}\par 
	\end{multicols}
	\caption{LV level forecasts present unique challenges. On the left are examples of a week's worth of demand from aggregations of 500 households (plot a) down to a single household (plot f). On the right is a illustration of the power law relationship of relative error as a function of feeder size.}
	\label{lvplots}
\end{figure*}

The larger focus of the literature on household-level forecasting, compared to the load at the more general LV level, can mainly be attributed to an increasing amount of smart meter data. LV network feeders consist on average of about 50 households \cite{Haben2019stl} and hence are typically less volatile than smart meter data. This is illustrated in the left hand plot in Fig.~\ref{lvplots}. This shows different aggregations of households from the Irish smart meter trial~\cite{Commission2012csm} from $500$ households (labelled (a) at the top) down to a single household at the bottom (labelled (f)). Beyond 100 households the profiles are relatively smooth but below this the data is increasingly irregular and volatile with varying degrees of spikiness.   

The challenges associated with modelling load at lower levels of aggregation are further illustrated in the right hand side plot in Fig. \ref{lvplots}. On the y-axis is the relative error (e.g. normalised root mean square error (RMSE) or mean absolute percentage error (MAPE)) and on the x-axis is the size of the LV feeder (e.g. its average daily demand, peak demand, number of consumers etc.). This illustrates a common feature of LV demand forecasts (alternatively forecasts of aggregates of household demand); there is often a power law relationship between the size of the feeder and the relative error \cite{mirowski2014dfi}. This means it becomes exponentially more difficult to accurately forecast smaller feeders (in terms of average demand or number of customers connected). The paper by Wang et al. \cite{Wang2018ros} is also one of the few reviews which mention LV level forecast, and also highlights the volatile nature of the associated demand behaviour. 

As shown in Haben et al. \cite{Haben2014ane}, traditional pointwise error measures such as RMSE and MAPE may not be appropriate (or informative) to describe the accuracy of forecasts of smart meter demand (i.e. individual households) due to the so-called \textit{double-penalty effect}. It is likely that this effect also holds for demand on small feeders or aggregations of small numbers of residential smart meters. However, this has not been investigated within the literature.  

In the emerging LV demand forecast literature, lack of real data means that either only a few substations are considered or the LV substations are artificially created from aggregations of smart meter data. Studying only a few substations limits the conclusions from any analysis. As figure \ref{lvplots} illustrates, LV networks consist of a wide variety of behaviours with the number of consumers connected being one of the largest indicators of demand accuracy. Without a large enough sample very few general conclusions can be established. In addition, LV networks are not simply the aggregation of individual households but consists of many different components, including street lights, cameras, and other street furniture. These connections may also be reconfigured over time (see for instance ~\cite{mirowski2014dfi}). Further, as shown in \cite{Haben2019stl} knowledge of the types of households is vital, for example, households with overnight storage heaters can produce dramatically different behaviours. 

To the authors' best knowledge, the paper by Haben et al. \cite{Haben2019stl} is the only one which considers forecasts of a relatively large number ($100$) of real feeders. This highlighted previously unknown results, such as the effect of a high proportion of overnight storage heaters and commercial customers, and the lack of influence of temperature on the forecast accuracy. It is vital that these results are replicated and further studies are developed to better understand the limitations and features of LV level forecasts.  

In short, LV level demand has unique features compared to medium (MV) and high voltage (HV) level demand:
\begin{itemize}
	\item Increased volatility due to lower aggregation of demand.
	\item Increased variety of demands with different feeders made up of different numbers and types of consumers.
	\item Less well understood explanatory variables
	\item An increased range and variety of applications and requirements for forecasts at the LV
	level.
\end{itemize}

As will be demonstrated in this review, these features will drive major differences in the techniques and methods which are applied to forecasting LV demand compared to what has traditionally been developed for HV or system level demand forecasting.

\subsection{Related Reviews}
\label{sec_related_rev}

Before proceeding with the core topics of this paper, we summarise the main recent reviews in the area of forecasting, smart meter forecasting and smart meter analytics. This will serve the purpose of 1) providing a high-level overview of forecasting from the system level to household level, 2) highlighting the need for this review and 3) surveying peer-reviewed methodologies for conducting a viable review, which we will emulate to provide consistency.   

Hong and Fan \cite{Tao2016pel} provide a tutorial review of probabilistic load forecasting. They give an outline of other reviews in the area, the main methodologies applied, applications, evaluation methods as well as future problems. In this list they include electric vehicles, wind and solar generation, and demand response, all topics very much within the remit of LV level.

A recent paper by Hong et al. \cite{Tao2020efa} focused on a review of smart meter data. They looked at a range of forecasting topics that are becoming more prominent (and will also feature in this review) including forecast combination (Section \ref{hybridSection}), hierarchical forecasting, and probabilistic forecasting (Section \ref{section_prob}). Further issues such as open data, the role of forecasting competitions, and publishing issues are also discussed. Wang et al. \cite{Wang2018ros} also perform a review of smart meter data analytics and highlights several open smart meter data sets. One of the aims of this review is to also highlight and identify many open data sets that researchers may use. To further support researchers, we are also publishing a list of relevant datasets with links to major papers, see Table \ref{tab:datasets}. We hope this review article, with the list of key papers and datasets, would provide a good starting point to anyone embarking on research in this important and evolving field of modelling LV load.

As with most reviews in other areas, both \cite{Tao2020efa} and \cite{Wang2018ros} use a Scopus search to identify the number of published papers and major journals that publish forecasting and smart meter research. 

A review on analysis of residential electricity consumption and applications of smart meter data is given in ~\cite{Yildiz2017rai}. This is a review/survey on analysis and applications of smart meter data, but lists some major forecast methods, common inputs to the forecasts, and gives an overview of the traditional and new error measures being applied. It contains also household level applications such as home energy management systems, anomaly detection, customer feedback and health care for the vulnerable. Again this review will consider all of these topics but within the wider LV context. 

As in other fields, deep learning approaches are getting more attention from researchers lately. An unpublished review of deep learning approaches can be found in~\cite{gasparin2019deep}. It is not limited to the LV-level but they explicitly compare deep learning approaches applied to household data.  In contrast, Yin et al.~\cite{yin2020aso} give a survey of the quite limited scope of deep learning approaches in the distribution network, presenting some examples of applications in load and renewable energy sources (RES) forecasting as well as fault detection.  

The above reviews do not investigate the low voltage distribution networks but instead consider smart meters \cite{Tao2020efa}, \cite{Yildiz2017rai}, or general forecasting for the higher voltage, system-wide or national level \cite{Tao2016pel}. As discussed in the previous Section \ref{sec_motivation_relatedrevs}, LV networks encompass a much wider range of problems and applications than associated with the above related reviews. LV network demand is much more volatile than higher voltage level demand and is extremely diverse. This is because they often serve different numbers and types of consumers, mixing residential, and small commercial consumers. As demonstrated in this review, LV demand forecast requirements can be very different to those used in more general load forecasting, requiring very different inputs, different methods and in some cases, very different error metrics. 

For smart meter forecasting, the challenges are very similar to LV forecasting. They both are typically very volatile and therefore may require similar techniques such as probabilistic forecasts to estimate their associated uncertainty properly. However, there are some key differences. Firstly, LV network demand is not simply the aggregation of individual consumers demand (e.g. from smart meters) \cite{Haben2019stl}, and the presence of street furniture and the diversity of sizes and types of LV networks gives them unique features (such as the power law in Figure \ref{lvplots}) which are not components of individual smart meter data. Secondly, the LV networks produce a whole range of applications that are not applicable at the end customer level. As will be explored in Section \ref{sec:LVLF-applications} this includes network control, microgrid energy trading, and flexibility applications. 

Given these specific requirements, this review focuses on the relatively underexplored area of low voltage level forecasts, their associated applications, the most significant openly available datasets, and challenges and recommendations going forward. It should be noted that although this paper is not focused on smart meter forecasting, smart meters have been included in this review as they illustrate some of the same challenges with LV level forecasting, in particular the increased volatility and diversity of time series.

\subsection{Literature Selection Methodology}
\label{sec:lit_selection_methodology}

As with the other relevant reviews summarised in Section \ref{sec_related_rev}, the search for relevant papers was conducted using the Scopus\footnote{\url{https://www.scopus.com/}} abstract and citation database that provided a user-friendly interface for refining and investigating our queries. The search query was applied to the article titles, abstracts and keywords of articles, and consisted of the following terms:
\newline
\texttt{
	(substation OR feeder OR "low voltage" OR "smart meter") AND (load OR electricity OR consumption) AND (forecast*)
},
\newline
where text in quotation marks indicates exact match and text followed by asterisk indicates words starting with this sub-string.
The search consists of the main keywords representing the level (LV, substation, etc.) and type (electricity, etc.) of forecasting. 

\begin{figure*}
	\begin{multicols}{2}
		\includegraphics[width=0.45\textwidth]{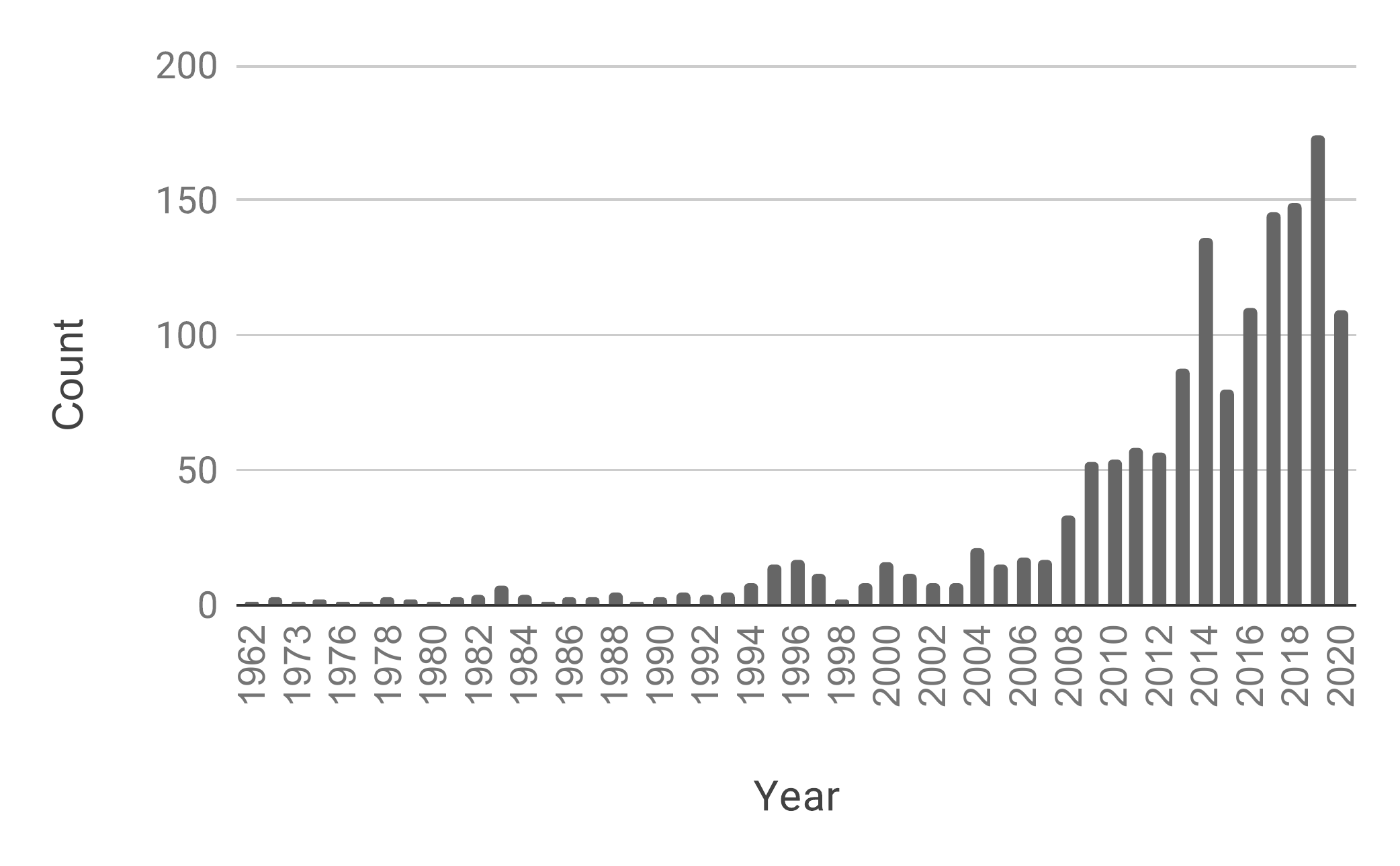}\par 
		\includegraphics[clip, trim=0cm 0cm 0cm -2.5cm,width=0.45\textwidth]{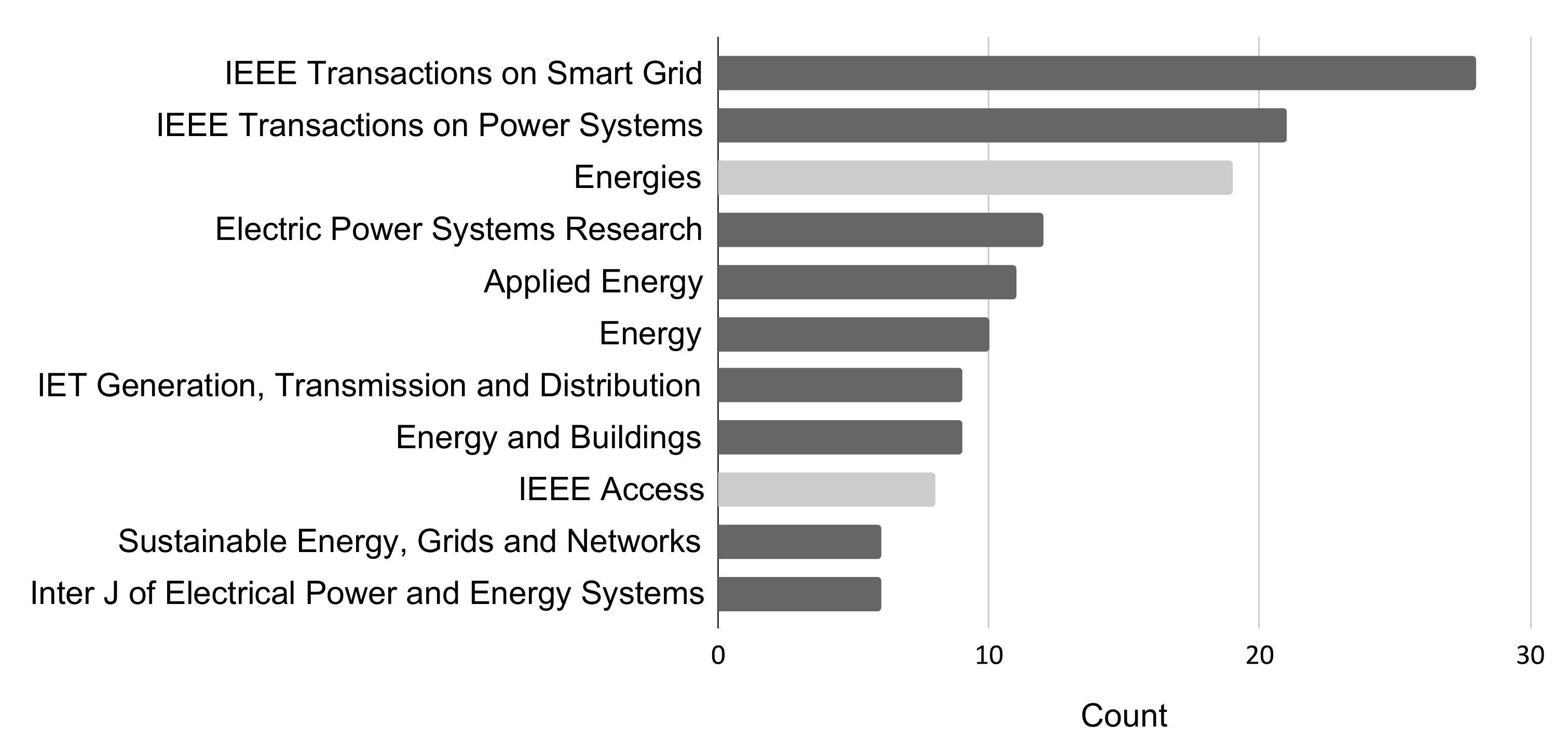}\par 
	\end{multicols}
	\caption{Publication count per year using the search terms as given in the text is shown on the left. On the right are top journals from the final set of articles. Open-access journals are depicted with a light grey colour}
	\label{fig:final_set_top_journals}
\end{figure*}

The final search before starting the reviewing process was conducted in August 2020 and it resulted in 1487 manuscripts. A breakdown of the number of papers (before filtering) is shown on a left panel in Figure \ref{fig:final_set_top_journals}, where we observe a proliferation of papers and high-interest post 2000. This is consistent with energy forecasting in general \cite{Tao2020efa} and smart meter analytics \cite{Wang2018ros}.

To manage the sheer number of papers, we proceeded with the following steps. We first removed any papers prior to the year 2000, which reduced the current paper count to 1362. Since we want to focus on peer-reviewed material and journals, we then filtered according to conference papers (conference paper and conference review) and others (article, article in press, books, book chapters, data paper, paper review). There are 807 conference and 555 non-conference items. For the non-conference we kept all 2020 and 2019 papers but only those older papers with five or more citations, as a proxy to impactful methods, which reduced the total to 423 non-conference papers (including 155 papers from 2019 and 2020). For the conference papers we were slightly stricter and kept only those with more than 20 citations while retaining all of those from 2020 resulting in 69 conference papers. In total, this resulted in 492 articles. 

A high level investigation of these papers identified several papers which were not about low voltage level load forecasting, could not be accessed because they were behind paywalls and could not be accessed by any others sources (including contacting the authors), or were not available in English. This provided a final list of 221 papers which were read and reviewed by the authors. It was found when reading some of these papers in detail that forecasts were not a topic of the manuscripts (typically consisting only to a possible application or only discussed conceptually). Hence the final number of reviewed papers is slightly smaller than 221. In addition, a few papers that are not connected to the keywords have been included in this review, because they tell the wider narrative such as the more general short term forecasting reviews discussed above (as well as a few methods' and dataset references). 

The most frequently occurring journals from the final list of reviewed papers are shown in Figure \ref{fig:final_set_top_journals}, where we find only two open-access journals: \textit{IEEE Access} and \textit{Energies}. \textit{IEEE Transactions on Smart Grid} and \textit{Power Systems}, respectively, are the most popular journals in the field, followed by \textit{Energies}.

\subsection{Structure of this Review}
% how this paper is structured
This review emulates several features from the previous reviews discussed in Section \ref{sec_related_rev} including investigating the methodologies, common explanatory variables, special topics like forecast combination and hierarchical forecasting but will be unique in several areas. Firstly, the applications are extended beyond the smart meter or system-level forecasting reviews, and discuss real-life planning, operations and control of LV networks. Since a major intention of this review is to encourage further research in LV level forecasting, this paper illustrates the opportunities and challenges and will establish an open community-driven list of known LV level open data sets to encourage further research and development. 

\begin{figure}
	\centering
	\includegraphics[width=0.60\linewidth]{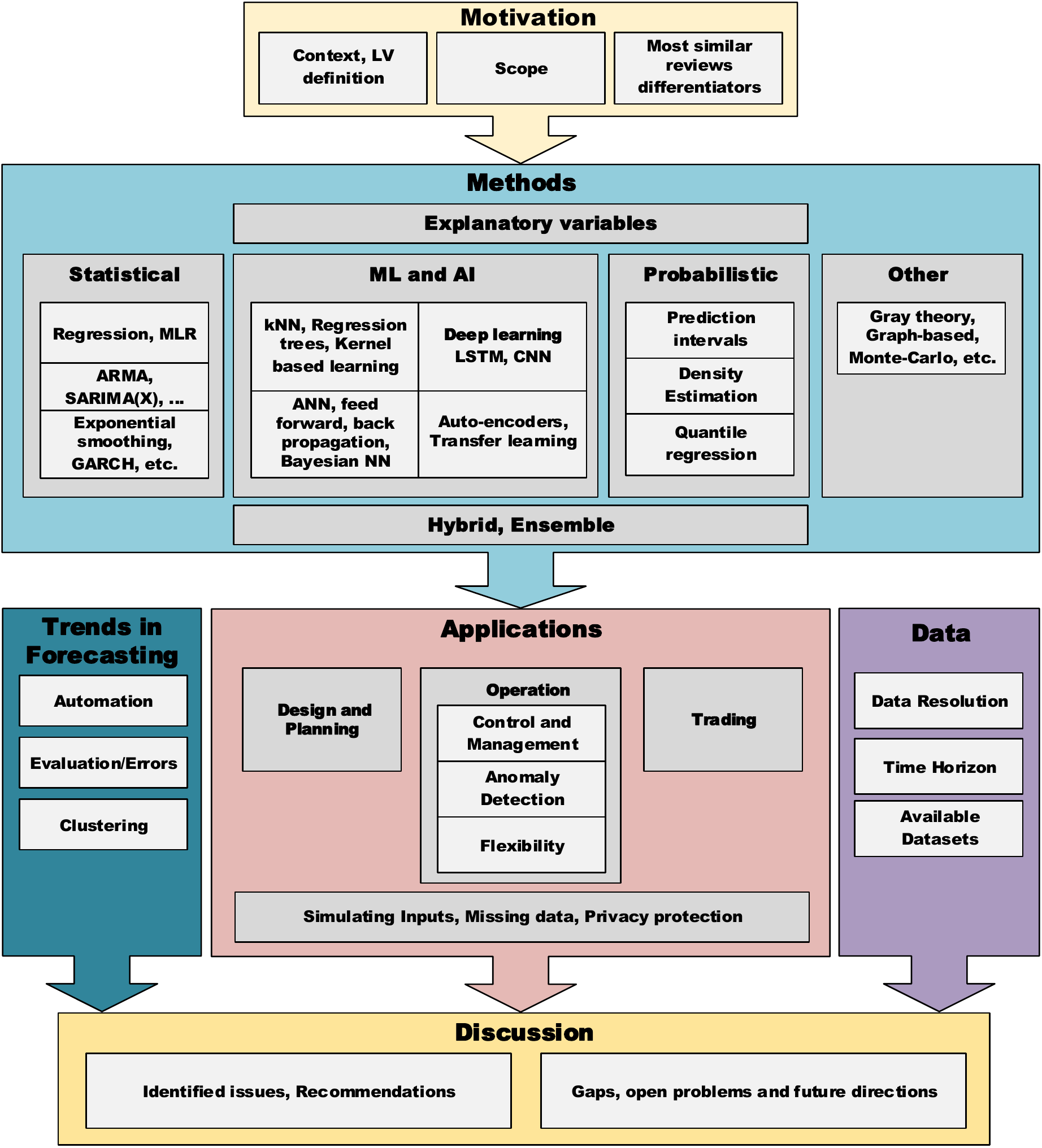}
	\caption{The structure of this review.}
	\label{fig:flowchart}
\end{figure}

Figure~\ref{fig:flowchart} provides an overview of the structure of the paper. 
Section \ref{sec_forecasting_method_main} provides a comprehensive overview of the main methods and techniques in LV level load forecasting, including point forecasts (both statistical and machine learning methods), probabilistic forecasts, combination approaches, as well as more esoteric methods. Section \ref{sec_trends_special_topics} presents some of the emerging trends and special topics. Section \ref{secdatasets} focuses on the LV datasets which are utilised in the reviewed papers, summarising some of the most commonly available datasets and their features. It links to the open data list. Section \ref{sec:LVLF-applications} focuses on the LV level applications which utilise forecasting. Finally, this review will conclude with a discussion in Section \ref{sec_discussion} identifying some of the major challenges as well as sharing some of the authors own views and recommendations for future research. Note, that while the Methods section is quite comprehensive, by structuring the work by their applied methods, Table~\ref{tab:methodoverview} allows a quick reference of the reviewed work by the aggregation level and forecasting horizon. Further, each of the sections is self-contained, so that a reader interested in trends found, applications and datasets identified, or the conclusions drawn, can skip ahead to the respective section.

\section{Low-Voltage Load Forecasting Methods}
\label{sec_forecasting_method_main}

This section will focus on the forecast models and techniques including the inputs and explanatory variables used in LV level forecasting methods found in the literature. It starts with a focus on the explanatory variables that have been used within the surveyed approaches. The remainder of this section structures the studied approaches by the type of method used. Table~\ref{tab:methodoverview} gives an overview of the surveyed methods of this section for reference by the aggregation level and the forecasting horizon. For the individual level, we distinguish between residential and commercial/industrial customers (or mixed if both are included or not specifically distinguished). For aggregate level we split these into those directly applied to substation data, or those which are the aggregation of individual consumers. We classify forecasting horizons up to a few hours as \emph{very short-term}, day-ahead and up to a few days as \emph{short-term}, \emph{medium term} from weeks to months and \emph{long-term} from a year ahead and up.

\begin{table}[]
	\scriptsize
	\caption{Overview of reviewed LV Load Forecasting Approaches by Aggregation Level and Forecasting Horizon.} \label{tab:methodoverview}
	\begin{tabular}{p{.06\linewidth}p{.08\linewidth}p{.08\linewidth}p{.46\linewidth}p{.29\linewidth}}
		\toprule
		\textbf{Level} & \textbf{Category} & \textbf{Horizon} & \textbf{Method} & \textbf{Reference} \\ 
		\midrule
		\multirow[t]{32}{*}{Individual} & \multirow[t]{11}{*}{Residential} & \multirow[t]{1}{*}{Very} & Statistical and time series Approaches & \cite{Ghofrani2011smb} \\
		&  & \multirow[t]{4}{*}{Short-term} & Machine Learning and other Artificial   Intelligence Approaches & \cite{Singh2018bdm, mehdipour2020slf, liu2019tsh, estebsari2020srl, kong2017str, Elvers2019spl, voss2018residential, voss2018adjusted} \\
		&  &  & Comparing Methods & \cite{Cordova2019cet, Cerquitelli2019esm, nawaz2019aaf, Humeau2013elf} \\
		&  &  & Probabilistic Forecasting & \cite{wang2019pil, yang2020bdl, wang2019pil} \\
		&  &  & Hybrid, Combination and Ensemble   Approaches & \cite{dong2016ahm, ai2019hpd} \\
		&  & \multirow[t]{5}{*}{Short-term} & Statistical and time series Approaches & \cite{kipping2016mad, Litjens2018aof, Arora2016fes, Chaouch2014cio, dinesh2020rpf, Nugraha2018ldp} \\
		&  &  & Machine Learning and other Artificial   Intelligence Approaches & \cite{Singh2018bdm, sousa2012atr, bessani2020mhv, shah2020stm, Elvers2019spl, Shi2017dlf, voss2018residential, Haben2014ane, voss2018adjusted} \\
		&  &  & Comparing Methods & \cite{Cordova2019cet, nawaz2019aaf, gajowniczek2014ste, Humeau2013elf} \\
		&  &  & Probabilistic Forecasting & \cite{Arora2016fes, Arora2016fes, pinto2017mpf, gerossier2018rda} \\
		&  &  & Hybrid, Combination and Ensemble   Approaches & \cite{dong2016ahm, Kiguchi2019pil} \\
		&  & Long-term & Machine Learning and other Artificial Intelligence Approaches & \cite{Singh2018bdm} \\
		\cmidrule(lr){2-5}
		& \multirow[t]{11}{*}{Mixed} & \multirow[t]{1}{*}{Very} & Machine Learning and other Artificial Intelligence Approaches & \cite{Hosein2017lfu, desilva2011ipc} \\
		&  & \multirow[t]{2}{*}{Short-term} & Comparing Methods & \cite{mirowski2014dfi} \\
		&  &  & Probabilistic Forecasting & \cite{Wang2019cpl, Yang2019del} \\
		&  & \multirow[t]{5}{*}{Short-term} & Statistical and time series Approaches & \cite{aprillia2019oda, Bhattacharyya2020sgd} \\
		&  &  & Machine Learning and other Artificial   Intelligence Approaches & \cite{moon2019aca, pati2020mfc, khan2020tee} \\
		&  &  & Comparing Methods & \cite{mirowski2014dfi} \\ 
		&  &  & Probabilistic Forecasting & \cite{Taieb2020hpf, chaouch2015rcq, Yang2019del, taieb2016fui} \\
		&  &  & Hybrid, Combination and Ensemble   Approaches & \cite{Amato2021fhr, taieb2016fui} \\
		&  & Med.-term & Hybrid, Combination and Ensemble Approaches & \cite{massidda2019smf} \\
		&  & \multirow[t]{2}{*}{Long-term} & Machine Learning and other Artificial Intelligence Approaches & \cite{fiot2018edf} \\
		&  &  & Hybrid, Combination and Ensemble   Approaches & \cite{massidda2019smf} \\
		\cmidrule(lr){2-5}
		& \multirow[t]{1}{*}{{Commercial/}} & \multirow[t]{1}{*}{Very} & Statistical and time series Approaches & \cite{ullah2018apm} \\
		& \multirow[t]{9}{*}{{Industrial}} & \multirow[t]{3}{*}{Short-term} & Machine Learning and other Artificial   Intelligence Approaches & \cite{Feng2020rda} \\
		&  &  & Comparing Methods & \cite{grolinger2016eff} \\
		&  &  & Other methods & \cite{xu2020ats} \\
		&  & \multirow[t]{4}{*}{Short-term} & Statistical and time series Approaches & \cite{Ding2015nms, ullah2018apm, Lee2013asm} \\
		&  &  & Machine Learning and other Artificial   Intelligence Approaches & \cite{ribeiro2018tls, petrosanu2019ddv, Jurado2015hme, jurado2017fir} \\
		&  &  & Hybrid, Combination and Ensemble   Approaches & \cite{ruiz-abellon2018lff, Jung2020bem} \\
		&  &  & Other methods & \cite{xu2020ats} \\
		&  & Med.-term & Machine Learning and other Artificial Intelligence Approaches & \cite{gao2020dla} \\
		&  & Long-term & Statistical and time series Approaches & \cite{tsekouras2006anl} \\
		\midrule
		\multirow[t]{21}{*}{Aggregate} & \multirow[t]{12}{*}{Substation} & \multirow[t]{1}{*}{Very} & Statistical and time series Approaches & \cite{Ghofrani2011smb, Bracale2013ABB} \\
		&  & \multirow[t]{3}{*}{Short-term} & Machine Learning and other Artificial   Intelligence Approaches & \cite{zhao2020onp, lourenco2012stl, wang2013acv} \\
		&  &  & Comparing Methods & \cite{idowu2016aml, mirowski2014dfi} \\
		&  &  & Hybrid, Combination and Ensemble   Approaches & \cite{Cao2020hed, sulandari2016fel, tajer2017lfv} \\
		&  & \multirow[t]{6}{*}{Short-term} & Statistical and time series Approaches & \cite{Haben2019stl, aprillia2019oda, vincenzo2020cml, borges2020etm, Goude2014lsa, Hayes2015acs, Haben2019stl, espinoza2005stl} \\
		&  &  & Machine Learning and other Artificial   Intelligence Approaches & \cite{bersani2006mol, bersani2006mol, sousa2012atr, ahmad2019dlf, ding2016nnb, stephen2020ngr, fidalgo2005lfp, naeem2020stl, zufferey2020psf, Abreu2018mlf, chen2018dfl} \\
		&  &  & Comparing Methods & \cite{idowu2016aml, mirowski2014dfi, Yunusov2018sss} \\
		&  &  & Probabilistic Forecasting & \cite{Haben2019stl, bikcora2018dfo, kodaira2020oes, zufferey2020psf, bikcora2018dfo} \\
		&  &  & Hybrid, Combination and Ensemble   Approaches & \cite{bogomolov2016ecp, hashim2019anl, Bennett2014flv, sun2015aea, grmanova2016iel, nespoli2020hdf, kodaira2020oes} \\
		&  &  & Other methods & \cite{mcqueen2004mcs} \\
		&  & \multirow[t]{2}{*}{Long-term} & Statistical and time series Approaches & \cite{Friedrich2014mfu, Goude2014lsa} \\
		&  &  & Machine Learning and other Artificial   Intelligence Approaches & \cite{Ye2019adb} \\
		\cmidrule(lr){2-5}
		& \multirow[t]{9}{*}{Aggregation} & \multirow[t]{1}{*}{Very} & Statistical and time series Approaches & \cite{perfumo2014mbe} \\
		&  & \multirow[t]{2}{*}{Short-term} & Machine Learning and other Artificial   Intelligence Approaches & \cite{lin2020ecp, Elvers2019spl, voss2018residential, voss2018adjusted} \\
		&  &  & Comparing Methods & \cite{Humeau2013elf} \\
		&  & \multirow{6}{*}{Short-term} & Statistical and time series Approaches & \cite{macmackin2019mad, dinesh2020rpf, Marinescu2013RED} \\
		&  &  & Machine Learning and other Artificial   Intelligence Approaches & \cite{moon2019aca, chen2019daa, sun2019ubd, wu2020ast, Elvers2019spl, voss2018residential, sousa2014slf, voss2018adjusted, konjic2005fis} \\
		&  &  & Comparing Methods & \cite{garulli2015mat, Humeau2013elf} \\
		&  &  & Probabilistic Forecasting & \cite{Taieb2020hpf, taieb2016fui} \\
		&  &  & Hybrid, Combination and Ensemble   Approaches & \cite{Laurinec2019due, Amato2021fhr, taieb2016fui} \\
		&  &  & Other methods & \cite{zhou2019pse} \\ \bottomrule
	\end{tabular}
\end{table}

\subsection{Explanatory Variables}
\label{subsec:expl_var}
Selecting the correct inputs (predictors/features) is almost as important as choosing the most appropriate models when developing a forecasting method. The selection of the most relevant set of explanatory variables is imperative for both the descriptive and predictive performance of a model. The literature review confirmed many of the most common choices of explanatory variables used in forecast models, as well as some novel features. While the impact of weather on the load at the national level is well studied, there is a need to understand the influence of weather predictions at the LV level, for both short- and medium-term forecasts. In this section, we review some of the most commonly used explanatory variables used for forecasting at the LV level.

The majority of models use some form of historical or lagged observations in the forecasts \cite{sousa2012atr, gajowniczek2014ste, Yunusov2018sss, jurado2017fir}. LV level demand is similar to higher voltage demand in that there is often daily, weekly and annual seasonalities that needs to be accommodated to generate forecasts. This also includes using calendar variables, \cite{Cerquitelli2019esm,ahmad2019dlf,moon2019aca, sousa2012atr}. In addition to lagged load values, Louren{\c{c}}o and Santos \cite{lourenco2012stl} used derivative terms (differences between the adjacent values), which was not found to be common in the literature review. 

\subsubsection{Meteorological variables}

By far the most common explanatory inputs are meteorological variables \cite{Jung2020bem}, in particular temperature \cite{hoverstad2015stl, kipping2016mad}. Temperature is also combined with other variables such as  humidity \cite{Marinescu2013RED}, solar irradiance \cite{dong2016ahm}, solar irradiance and wind speed \cite{larsen2017dre}, or dry-bulb temperature, dew point, precipitation rate, wind chill and humidity index \cite{lusis2017str}. One paper Zhou et al. \cite{zhou2019pse} combines temperature, precipitation, pressure and wind speed. Additionally, the weather variables can be combined to produce joint variables as in \cite{ding2016nnb} which combines lagged, calendar and temperature and the cross product of these features for capturing non-linearities.
If the raw meteorological variables or their lagged counterparts are not used as inputs then derived features are used instead, for example cooling degree days (CDD) and heating degree days (HDD), which measure temperature exceedances from a given threshold value. Commonly an exponentially smoothed version of the temperature is used as in \cite{larsen2017dre}. This can help take into account the delayed effect of temperature on demand.

The interaction between explanatory variables or their effect (whether positive or negative) in forecast models has only been touched upon in a few of the papers considered in this review. Lusis et al. \cite{lusis2017str} find that calendar effects have less predictive power when used with the weather, daily and weekly seasonality. Fidalgo and Lopez \cite{fidalgo2005lfp} state that experiments with temperature features did not strongly affect the forecast accuracy and were discarded. Furthermore, Haben et al. \cite{Haben2019stl} consider several models with and without temperature data (both forecast and actuals) and they found no, or negative effect on the short term forecasts accuracy for both point and probabilistic forecasts. In contrast, Bennett et al. \cite{Bennett2014flv} incorporate temperature and relative humidity (RH) for generating point day-ahead LV transformer level forecasts, and report that temperature accounted for about half the variation in the load. Transformed values of the temperature (squared temperature) and interaction effects (RH with temperature) were also considered.

This highlights a common feature of the models which use meteorological explanatory variables. Ideally, in practice, the predicted load should only rely on using the predictions of weather variables as explanatory variables (so-called \textit{ex-ante forecasts}), which are preferably obtained from several weather stations in a geographic neighbourhood. However, a vast majority of the reviewed papers employ the actual weather observations (\textit{ex-post forecasts}). It can be thus expected that the corresponding forecast accuracy is over-optimistic. To the best of our knowledge, other than Haben et al. \cite{Haben2019stl} there are only two other papers found in this review using weather forecast inputs \cite{prakash2018rbe, nespoli2020hdf}.  In \cite{nespoli2020hdf}, the authors considered a large number of different variables including temperature, global horizontal and normal irradiance, relative humidity, pressure and wind speed and direction. In \cite{prakash2018rbe} the authors developed a Gaussian Process regression-based 10-minute ahead forecast for the aggregated consumption of ten buildings using both actual and predicted temperature. In this case, using predicted temperature produces smaller errors than using actual for the longer-term forecasts. Although only a handful of papers include weather predictions as model inputs for LV load forecasting \cite{Haben2019stl, prakash2018rbe, nespoli2020hdf}, it is worth noting that these papers only employ point estimates and not weather ensemble predictions, thereby ignoring the underlying uncertainty.

Since these papers consider LV level, they require more local but unknown weather inputs. The Irish CER dataset~\cite{Commission2012csm} is a common dataset used despite the locations of the households being unknown. Hence many authors use an average over the entire country or use the weather of a major city such as Dublin as representative\cite{yang2020bdl}.

\subsubsection{Econometric variables}

Econometric variables or information about the types and number of consumers on an LV network are very common inputs \cite{Bunnoon2013mcc}. These variables are typically used for longer-term forecast models, for example, \cite{tsekouras2006anl} uses a whole host of information about manufacturing, number of consumers and gross national product to produce mid-term length forecasts. Additional features from surveys, such as the number of residents, social class and electricity use for heating and cooling are used in \cite{bessani2020mhv}. In a similar way, a range of different variables were considered in \cite{Kiguchi2019pil,kipping2016mad}, including gender, age group, social class, and the number of other residents.  The number of customers on an LV substation and the monthly energy consumption is used in \cite{konjic2005fis}. In \cite{idowu2016aml}, the authors used the substation internal state in addition to lagged load, temperature and temporal features, which proved to be significant in improving the forecast.

Inputs from other monitored customers are also common. For forecasting (and imputing) the load of one substation~\cite{borges2020etm} uses neighbouring substation's data to improve forecasts.  Ziekow et al.~\cite{ziekow2013tpo} analyze how the availability of submetered data impacts very short term forecasts (15-minute and 1-hour-ahead). They generally find improvements varied among three households from 5\% to above 30\%. They also find that higher resolution produces greater accuracy improvements in very short term forecasts (15 min ahead).  

\subsubsection{Novel variables}
\label{subsec:novel_var}
Besides weather, calendar, and econometric explanatory variables, other explanatory variables are also employed for modelling LV load. The following is a list of some further examples of more novel explanatory variables which were identified in the literature review:
\begin{itemize}
	\item For forecasting the change in demand side response, in~\cite{garulli2015mat}, active  demand (requested change in demand) is used as an explanatory variable.
	\item For sites with photovoltaics (PV), a net load forecast in~\cite{chu2017nlf} was produced using the red-blue ratios in sky images to derive features for inputs to the model. 
	\item In \cite{Cordova2019cet} electricity data is combined with transport data.
	\item To produce forecasts for an event-organizing venue, event type, schedule, day-of-the-year, and seating configurations are used as inputs in~\cite{grolinger2016eff}.
	\item In~\cite{bogomolov2016ecp} electricity data is combined with mobile usage data. Interestingly, the authors do not use any energy data as predictors. Also, although daily and weekly seasonalities were present in the data and coded into the feature space, these features are not present in the final feature list. Finally, the three different areas that have different demand patterns are not coded into the features either. 
	\item  Forecasting in terms of the customer baseline load (CBL), i.e., the consumption level that a customer would have have consumed in the absence of any demand response program is considered in ~\cite{pati2020mfc} using the CER Irish smart meter data.
\end{itemize}

\subsection{Statistical and Time Series Approaches}  
\label{secStatandTS}
Statistical and time series approaches are one of the most commonly used models for short term load forecasting. The majority of them are linear in their parameters and include multiple linear regression (MLR), exponential smoothing and traditional time series approaches such as autoregressive and moving averages models (ARIMA, AR, etc.). However, many nonlinear approaches such as nonlinear regression, and generalised additive models have become more popular in recent years. 

\subsubsection{Linear Regression}
Regression estimates the relationship between one or more predictors (explanatory variables)  and the variable one wants to predict (dependent variable), in our case, the load at LV level. Multiple linear regression (MLR) assumes that the relationship between explanatory and dependent variables can be adequately modelled using a linear modelling approach (i.e., no deviations from the Gauss-Markov assumptions), whereby the model parameters are typically estimated based on the minimization of residual sum of squared errors. Some of the simplest linear models are applied by \cite{Litjens2018aof}, including seasonal persistence models and simple averages. Although simple in their formulation, such models can serve as sophisticated benchmarks during the out-of-sample validation of more sophisticated models \cite{Haben2019stl}.

Linear models are popular because they are easy to interpret and solve, and despite the constraints on linear coefficients, they can model a wide diversity of behaviours, including non-smooth and nonlinear relationships. In \cite{macmackin2019mad}, an MLR model is used for forecasting aggregate smart meter data from a utility in Canada. Dummy variables are included for types-of-day, and change points define heating and cooling degree days/hours. Haben et al.~\cite{Haben2019stl} also consider dummy variables for modelling types of the hour and day, and include Fourier components to model annual seasonalities. In \cite{Ding2015nms}  day-ahead forecasts are generated for a medium voltage/low voltage (MV/LV) substation, using a regression-based model with Fourier components, however, the results are compared with only a na\"{\i}ve benchmark based on a random walk model. The authors in \cite{aprillia2019oda} introduce an MLR method integrated with discrete wavelets transform, to produce a day ahead hourly load forecast at both the system and end-users levels. This model uses both the weather and lagged features and is benchmarked against multiple methods including standard MLR, Artificial Neural Networks (ANN), Autoregressive moving average model with external variable (ARMAX), and Support vector regression (SVR). 

Linear models are often coupled with meteorological data. Borges et al.~\cite{borges2020etm} use linear models with different subsets of features for short-term forecasting and missing data imputation for substation data. Their model uses historic values, meteorological data and neighbouring substation data. Bhattacharyya and Bhattacharyya (2020) \cite{Bhattacharyya2020sgd} forecast hourly load profiles for demand-side management using an MLR model with temperature, heating and cooling degree days as inputs. 
Haben et al.~\cite{Haben2019stl} incorporated temperature observations and forecasts into their linear models via a third-degree polynomial. In \cite{kipping2016mad} two MLR load forecasting models are developed, based on daily and hourly mean values of outdoor temperature, for a household in Norway. 
The authors report that outdoor temperature, dwelling group, floor space, and number of residents are the most important variables required for modeling hourly electricity consumption in dwellings with electric heating. Also, the model with HDD achieve slightly higher accuracy. 

In \cite{vincenzo2020cml}, the authors study a micro-grid that consists of a block of three offices, PV generation, a storage system and three smart charging electric-vehicles (EV) stations. The idea is to find the optimal operation of the integrated energy systems in order to reduce the peak. 

To forecast the building consumption, an MLR is used and feature selection is performed using a genetic algorithm 
on lagged load values, e.g. 1-3 days before, and weather variables.

Although mostly solved by least-squares estimation the linear state model (linear regression) can also be solved using a Kalman filter as in \cite{Ghofrani2011smb} where the household demand is split into a deterministic part (modelled by a 10-degree polynomial) and the remaining random part modelled as a linear Gaussian-Markov process. Another use of a Kalman filter was presented in  \cite{perfumo2014mbe} to generate probabilistic hour ahead forecasts of the aggregated demand of 70 households.

\subsubsection{Time Series Models}
\label{sec:tsmodel}
Time series methods, including autoregressive (AR), moving averages (MA), autoregressive integrated moving average (ARIMA), and those with exogenous inputs (ARX, ARIMAX etc.) are commonly used for demand forecasting. However, despite the daily and weekly seasonality of demand time series the seasonal version of these models (e.g. SARIMA) appear to be used less.

In \cite{Marinescu2013RED} a few models are compared, including ANN, fuzzy logic and wavelet neural networks, with the ARIMA model performing the best. In \cite{Nugraha2018ldp}, an ARIMA model is used for short-term (one hour-ahead and one day-ahead) forecasting of smart electricity meter data for building energy management systems. For the shorter-term forecasts, they use an online ARIMA. An ARIMA model is also used in \cite{Lee2013asm} for forecasting electricity for public school buildings, using historical load and temperature, to assist building management systems. They use variable base degree day/hour regression models combined with ARIMA to access energy efficiency, predict energy consumption, and detect energy usage anomaly.

While ARIMA models are often used to forecast the time series, in \cite{Bracale2013ABB} an AR(1) process is deployed to forecast the mean parameter of a Gaussian distribution of a Bayesian model. The standard deviation is similarly updated but with a Gamma distribution assumed.

Espinoza et al. \cite{espinoza2005stl} use a unified modelling framework based on periodic autoregressive (PAR) models for forecasting and clustering load profiles using data from 245 HV and LV substations from the Belgian National Grid Operator Elia.

\subsubsection{Nonlinear regression, exponential smoothing, and other models}

Although standard linear regression and time series models have been successfully applied to demand forecasts at all levels of the LV network, nonlinear regression models are also considered. Nonlinear models are very versatile but may be more prone to overfitting. This can be mitigated with cross-validation and other specific methods, such as model penalties in generalised additive models to tune smoothness.   

Hayes et al. \cite{Hayes2015acs} use a nonlinear autoregressive exogenous (NARX) model to forecast smart meter loads, which are shown to outperform both traditional ARX models and a neural network model. A nonlinear multivariate regression is used by Tsekouras et al.  \cite{tsekouras2006anl}, they select models based on testing multiple combinations of nonlinear functions to produce a medium-term forecast. The authors in \cite{Friedrich2014mfu} propose linear and multiplicative, nonlinear trend regression models to generate medium-term load forecasts (up to a year ahead), using substation-level data and weather data. The novelty of this work is regarding an estimate of the fraction of electricity load that can be attributed to cooling, which is a major consumption factor in the United Arab Emirates. A feature of this model is the use of dummy variables to differentiate Fridays/Holidays, Saturdays and Ramadan days from workdays.

Nonparametric and semi-parametric approaches can be beneficial in that they make fewer assumptions regarding the underlying data generating process than the fully parametric models. Chaouch \cite{Chaouch2014cio} uses a functional wavelet-kernel approach with clustering to forecast 2000 households from the Irish smart meter trial~\cite{Commission2012csm}. Goude et al. \cite{Goude2014lsa} generate short- and medium-term load forecasts for 2200 distribution network substations in France using a semi-parametric additive model. Although additive models have been quite successful at higher voltages (e.g. winning the GEFCOM 2014 challenge) to the authors' knowledge this is the first example of them being applied to low voltage demand. The authors state that modelling the middle term trend is quite challenging, and they did not have access to relevant commercial or sociological variables for modelling the trend in this study.

Although relatively simple linear models, exponential smoothing methods, which put more weight on recent past loads than older data, have been shown to be quite powerful. The double seasonal exponential smoothing method, also know as Holt-Winters-Taylor (HWT),  was one of the best performing in generating both point and probabilistic estimates against a multitude of methods as shown in \cite{Haben2019stl} and \cite{Arora2016fes}. 

In \cite{dinesh2020rpf} the authors use a bottom-up approach to predict load at the household and micro-grid level from individual appliances. The novelty of this paper is that statistical relationships between appliances are modelled (time-of-day probabilities and state duration probabilities). Ullah et al. \cite{ullah2018apm} also use a statistical approach by employing  Hidden Markov Models (HMM) to predict the consumption of residential buildings in Korea. The energy data for each floor is transformed into a floor occupancy sequence which are the observed values of the HMMs.

\subsection{Machine Learning and other Artificial Intelligence Approaches} % GG, MV
\label{sec:machine-learning}

The advances in the machine learning domain are also reflected in the LV load forecasting domain. ANN were introduced in the 90s and have been a popular method ever since. However, other methods based on nearest neighbours, regression trees, and kernels have also been popular over the past decades. With the recent advances of deep learning in other areas, many deep approaches have been proposed for LV load forecasting. 

\subsubsection{Nearest Neighbors}

The k-nearest neighbours (kNN) algorithm can be used for nonparametric regression. It is a simple method that often functions as a robust benchmark and is therefore included in many comparative studies (see Section~\ref{sec:comparing_methods}). Voss et al.~\cite{voss2018adjusted} show that at the household-level where load profiles are somewhat intermittent, that kNN can be adjusted to minimise the adjusted error measures proposed in ~\cite{Haben2014ane} (see Section~\ref{sec:application_specific}). As shown in Section~\ref{section_prob}, it can also be used for probabilistic forecasting by optimally weighting the k nearest neighbors~\cite{zufferey2020psf}. De Silva et al.~\cite{desilva2011ipc} propose an online learning algorithm, called Incremental Pattern Characterization Learning, based on a data stream mining algorithm and is related to nearest-neighbour methods. Considering the load forecasting problem as an incremental learning problem, it mitigates some weaknesses of static machine learning (ML) approaches by better handling concept shifts in the load data (e.g. expanding grid, changing customers or infrastructure, etc.).

\subsubsection{Regression Trees} \label{sec:trees_and_forests}

Regression trees (RT) are a form of decision trees where the target variable takes a continuous set of values. Compared to other machine learning models, regression trees can still be interpretable if they maintain a reasonable size. Chen et al.~\cite{chen2018dfl} investigate day-ahead peak load forecasting at the feeder-level using a parametric version of RT, Bayesian additive regression trees (BART), and report that BART generates more accurate point forecasts compared with linear regression, SVMs, and composite kernel methods based on Gaussian process regression.

\subsubsection{Kernel-based Learning Approaches}

A Gaussian process (GP) is a stochastic process based on lazy (just-in-time) learning and a kernel function that can also be used for probabilistic estimates based on a Gaussian distribution. Lourenco et al.~\cite{lourenco2012stl} use GP regression for an hour ahead load forecasting of three substations using the variance for estimating forecast confidence intervals. In \cite{Ye2019adb} a hierarchical approach for spatial load forecasting is proposed, to assist medium and low-voltage planning. They employ load, geographic, meteorological, economic and spatial data explanatory variables in their modelling. Specifically, they use non-parametric kernel density estimation (KDE) and adaptive k-means to generate aggregate spatial load densities, while stacked auto-encoders are used for forecasting the spatial load densities. Fiot and Dinuzzo \cite{fiot2018edf} generate forecasts of load at the smart meter level using kernel-based multi-task learning that considers the relationships between multiple demand profiles using the Irish CER data~\cite{Commission2012csm}. They demonstrate that kernels with multiplicative structure result in better forecast accuracy compared to additive structure kernels.

Support Vector Machines (SVM) is a robust, non-parametric method that is also based on kernel functions and has been popular, especially for smaller machine learning problems. Support Vector Regression (SVR) is the application of SVM to regression problems. Sousa et al.~\cite{sousa2014slf} propose an approach based on SVR for day-head residential load forecasting using simulated annealing for hyper-parameter search. The SVR model takes a representative load profile for a cluster of similar load profiles as input. The SVR with automated hyper-parameters can outperform a manually tuned ANN. Wang et al.~\cite{wang2013acv} propose a three-stage process using SVR at the feeder level, finding SVR outperforms ANN.

\subsubsection{Artificial Neural Networks}

Artificial Neural Networks (ANN) have attracted much attention in machine learning and they have been very popular in load forecasting and time series forecasting in general. The most basic form is a feed-forward neural network that has one or multiple hidden layers and is trained using the back-propagation algorithm and (stochastic) gradient descent. 

\paragraph{Feed-forward ANN} This basic ANN may be referred to as vanilla ANN or multi-layer perceptron (MLP) and is a popular benchmark for other approaches (see Section~\ref{sec:comparing_methods}) and the basis for more complex neural network architecture (see Section~\ref{sec:deep_learning}). Such a feed-forward ANN has, for instance, been used in \cite{sousa2012atr} for day-ahead forecasting of LV customers and substations in Portugal. They use a single hidden layer and an output layer with 24 nodes for each value of the hourly day-ahead profile. The authors compare different architectures, numbers of hidden nodes and activation functions. Similarly, Moon et al.~\cite{moon2019aca} explore day-ahead forecasting in buildings and groups of buildings in South Korea. They find that Scaled Exponential Linear Units outperforms other activation functions like the popular ReLU and that five hidden layers exhibited the best average performance. Pati et al.~\cite{pati2020mfc} compare ANN with simple moving average and exponential moving average methods for the Customer Baseline load (CBL) problem. 

A simple statistical benchmark method MidXofY (excluding Z lowest and Z highest load days to predict CBL)  results in the highest prediction accuracy for most of the consumers, followed by exponential moving average and an ANN method. In \cite{fidalgo2005lfp}, the authors develop one-hour to one-week load forecasts of 800 primary substation feeders and 200 transformers in Portugal. Their methodology is split into two parts. The first part is forecasting power, reactive power and current for a number of substations and feeders using ANN. The second part is about identifying anomalous events, such as historical data bugs, holidays, consumption price modifications, and re-training the ANN models. They find that retraining the ANN once the performance for a group of feeders is degrading for a period of four weeks, can make the forecasts more robust. Besides forecasting the load, ANN can also be used for imputing missing values of electricity consumption at the LV level.

\paragraph{Feature Transformations} While many authors use features directly related to load demand, others explore more advanced feature transformations. 
For instance, Bersani et al.~
\cite{bersani2006mol} explore Wavelet transforms as input for an ANN for day-ahead and intraday scheduling of LV transformers. Ding et al.~\cite{ding2016nnb} use ANNs for short-term forecasting of two MV/LV substations on the French distribution network. They split the load profile into two parts (i) daily average power and (ii) intraday power variation. Each component is forecast separately with different features via an ANN with a single hidden layer, but different model complexities (i.e. number of hidden neurons). ANN outperforms a naive and a linear regression model.

\paragraph{Beyond Back-Propagation} Besides regular back-propagation, also variants like Broyden–Fletcher–Goldfarb–Shannon algorithm, Quasi-Newton back-propagation and a One-step secant for gradient descent in back-propagation neural networks are explored, for instance at the district level (cf. \cite{ahmad2019dlf}).
Shah et al.~\cite{shah2020stm} propose an ANN that is optimised using particle swarm optimization (PSO) instead of back-propagation. They compare it to regular ANN and also long short-term memory (LSTM) and find that the PSO ANN substantially outperforms the other two. Extreme Learning Machines (ELM), introduced initially as randomized neural networks~\cite{schmidt1992ffn}, are a variant of ANN that use the Moore-Penrose generalized inverse to determine the weights instead of gradient-based back-propagation. Stephen et al.~\cite{stephen2020ngr} compare ELM with a standard multivariate Gaussian and a Gaussian Copula for probabilistic forecasting of LV feeders, showing that ELM outperforms the benchmark. Similarly, Zhao et al.~\cite{zhao2020onp} use ELM to produce probabilistic forecasts, focusing on prediction intervals.

\paragraph{Bayesian neural networks}
Bayesian neural network (BNN) are extending standard ANN with posterior inference. Bessani et al~\cite{bessani2020mhv} use a BNN for day-ahead forecasting. They train a global model, i.e., one that uses all available data of multiple households instead of fitting one local model per household. This approach outperforms a regular ANN benchmark as well as a HMM. Bayesian Networks are also used in \cite{Singh2018bdm} to predict energy consumption at the household level from individual appliances.

\subsubsection{Deep Neural Networks} \label{sec:deep_learning}

With the advancements in the area of deep learning, deep methods are also gaining relevance in the LV forecasting domain. Initially, recurrent approaches using long short-term memory (LSTM) cells were most popular. Compared to vanilla neurons, LSTM cells contain several gates, each with several trainable parameters, increasing the overall number of parameters to estimate from the training data. 

\paragraph{Long Short-term Memory} Kong et al.\cite{kong2017str} compare LSTM to regular ANN, ELM, kNN, and a na{\"i}ve model and find that it outperforms the other approaches in one-step-ahead forecasting (30 minutes) on 69 Australian households at the individual level. In contrast, at the aggregate level, LSTM perform similarly to regular ANN. Similarly, Mehdipour et al.~\cite{mehdipour2020slf} compare LSTM to SVR, Gradient Boosted Regression Trees (GBRT), and ANN at the household-level and also find that LSTM can consistently outperform the other models. Petroșanu et al.~\cite{petrosanu2019ddv} find that the combination of an LSTM with an NARX Feedback Neural Network can produce improvements over either alone. In \cite{lin2020ecp} an LSTM with an attention mechanism is proposed for short-term load forecasting in the China Southern Power Grid and four types of aggregated load; ‘Residential’, ‘Large Industrial electricity’, ‘Business’ and ‘Agricultural’. Particularly, the attention mechanism is used to assign weight coefficients to the input sequence data so that the specific features can be accurately extracted. The authors compare the proposed method to five benchmarks, an LSTM without attention, Holt-Winters, ARIMA, SVM and ANN and find that it outperforms each of the state-of-the-art methods.

\paragraph{Addressing Overfitting} Several studies find that due to the comparatively large number of trainable parameters, LSTM's trained for specific time series (e.g., a household or feeder) may be prone to overfitting. Mehdipour et al.~\cite{mehdipour2020slf} compare how different machine learning models generalise by assessing one-hour ahead forecasts on household data they have not trained on. For that, they try including 11, 22, 33, and 45 households in the training set and find that LSTM's improve as they have access to more data, while the other models' performances remain similar or deteriorate. The approach by Shi et al.~\cite{Shi2017dlf} uses this property of LSTM and finds that training a model on a pool of similar households improves over only training a local model. They find that this allows for deeper architectures and more accurate forecasts. Sun et al.~\cite{sun2019ubd} propose another approach based on incorporating deep Bayesian learning (DBL) into LSTM to avoid overfitting. DBL avoids overfitting by imposing a prior on hidden units or neural network parameters. Another direction is taken by Wu et al.~\cite{wu2020ast}. They use a cell related to the LSTM cell, the gated-recurrent unit (GRU). It has fewer parameters than the LSTM and can hence also avoid issues with overfitting. Further, they employ early-stopping as regularisation as a way to prevent overfitting.

\paragraph{Convolutional Neural Networks} In addition to approaches based on recurrent models like LSTM and GRU, methods based on convolutional neural network (CNN) layers have been proposed. Voss et al.~\cite{voss2018residential} apply a time series specific CNN, the WaveNET architecture based on causal and dilated convolutions at the household-level and show that it can outperform more straightforward approaches like MLR and ANN. Besides time series specific convolutions, two- and three-dimensional encodings of time series into images have also been applied. Estebsari and Rajabi~\cite{estebsari2020srl} compare three different image encodings based on recurrence plots, the Gramian angular field and the Markov transition field, finding that recurrence plots work well for encoding the time series as input for the convolutional layers. Elvers et al.~\cite{Elvers2019spl} utilise the daily seasonality and arrange the time series by aligning the daily load profiles row-wise in a 2D input matrix for regular CNN layers. The proposed model minimises the pinball loss for quantile regression and outperforms a quantile regression approach in intraday and day-ahead forecasting at the household-level and different aggregations.

\paragraph{Auto-encoders for Feature Representation} The previous applications used either purely LSTM or CNN layers directly to produce a forecast. Literature has investigated the combination of these different layers. Chen et al.\cite{chen2019daa} propose a multi-step load forecasting approach to aggregate load in the LV grid. After clustering similar customers, they use an auto-encoder (AE) based on convolutional layers for dimensional reduction. LSTM are then used to create a forecast per cluster in the encoded latent space, which is then decoded for the final forecast. They find that three neurons in the latent space are enough to represent the 48 dimensions of the load profile for the best forecasting results. Similarly, Khan et al.~\cite{khan2020tee} use a convolutional layer to extract spatial features, which are fed into an LSTM-AE to generate encoded sequences. A final dense layer is used for energy prediction. Here, the auto-encoder is used to tackle the issue that LSTM fails to learn temporal dependencies from one sequence to another. The method is applied for forecasting Korean residential and commercial building load data. Liu et al.~\cite{liu2019tsh} also find that a sparse encoding network can improve the forecast for an LSTM at the household-level. Naeem et al.~\cite{naeem2020stl} develop a day-ahead load forecast of an Australian network-grid using an Ensemble Empirical Mode Decomposition (EEMD) to decompose the signal into Intrinsic Mode Functions (IMF) and residuals. These modes and residuals are passed onto a Denoising Auto Encoder (DAE) for feature extraction. The output from the DAE is passed on to a CNN. The proposed method outperforms a deep belief network and an empirical mode decomposition mixed kernel-based ELM benchmark. % 

\paragraph{Comparing Deep Models} Hosein and Hosein~\cite{Hosein2017lfu} compare different neural architectures (regular ANN, ANN with fully connected AE, recurrent neural network (RNN) and LSTM). They find that an ANN combined with an AE is the best deep architecture and outperforms other traditional models such as weighted moving average, linear regression, regression trees, SVR, and MLP.

\paragraph{Transfer Learning} As mentioned above, deep architectures can benefit from using data not only from the local level but also from other (similar) buildings or households to improve accuracy. This property of deep learning models can be used in the cold start setting when not much data is yet available for a new building or a building where metering infrastructure is just deployed. For that transfer learning can be utilised. For instance, Gao et al.~\cite{gao2020dla} apply transfer learning from buildings with similar (i) environmental variables (close proximity) and (ii) functional type, e.g., office buildings. 
Their approach makes further use of pre-training and data augmentation. They find that their transfer approach can increase accuracy by around 20\% in settings with few available data. 
Similarly, Ribeiro et al.~\cite{ribeiro2018tls} introduce their transfer learning approach Hephaestus, a cross-building energy prediction method based on transfer learning with seasonal and trend adjustment. It is an inductive transfer learning method that is independent of the specific data-driven algorithm. 

\subsubsection{Other Artificial Intelligence Methods}
The aforementioned machine learning models are based on supervised machine learning, sometimes in combination with unsupervised approaches (see Section~\ref{sec:clustering}). Another class of machine learning algorithms is reinforcement learning. Feng et al.~\cite{ Feng2020rda} use Q-learning for model selection out of a pool of candidate ML models (ANN, SVR, GBM, RF) and for probabilistic forecasting from candidate distributions (Gaussian, Gamma, T, Laplace). Q-learning is used to choose the best model in each time-step, based on performance in a sliding window of data. They find that their ensemble approach outperforms the best individual models. 

By far, the most popular and successful artificial intelligence (AI)-based approaches, as introduced in the former section, are based on only a sub-section of artificial intelligence, i.e., machine learning, the sub-field that deals with algorithms that learn from data. Some approaches from fuzzy logic, another subfield of AI, are proposed or combined with ML approaches. Konjic et al.~\cite{konjic2005fis} use a so-called Takagi-Sugeno Fuzzy inference system for forecasting different classes of LV substations. Jurado et al.~\cite{Jurado2015hme} propose a forecasting method based on Fuzzy Inductive Reasoning (FIR) at the building level. They show based on three buildings that it outperforms ANN, RF and ARIMA. In \cite{jurado2017fir}, the authors introduce an improved version of the FIR algorithm, called flexible FIR, for studying missing values occurring in the training and test sets of load data. The method is applied to eight buildings in Catalonia (Spain), and it is demonstrated that increasing the percentage of missing values increases the modelling error. However, further details regarding hyperparameter tuning could have been provided. 

The work of Abreu et al.~\cite{Abreu2018mlf} combines ML with symbolic approaches by employing an ANN from fuzzy-adaptive resonance theory (which they refer to as fuzzy ARTMAP neural networks) for load forecasting, while considering different hierarchies of the distribution system. Their modelling comprises two components: global load forecasting (considering the sum of all loads on the system), and multimodal load forecasting (focusing on substations, transformers, and feeders).

\subsection{Comparing Methods} 
\label{sec:comparing_methods}

Many authors perform comparative studies where the focus is not on a single method, but on several different statistical and machine learning methods. Humeau et al.~\cite{Humeau2013elf} compare MLR, SVR and ANN at different aggregations of households and find that MLR works best at the household level and SVR performs best at the aggregate level (after 32 households). Given that both ANN and SVR are popular, others have also compared them. In \cite{gajowniczek2014ste}, the authors develop two day-ahead load forecasts at hourly resolution at the household level in Warsaw, using ANN and SVR. Particularly, the authors train a single ANN with 24 output nodes and 24 SVR models one for each hour of the horizon. However, the paper does not mention any comparison of the proposed method to any benchmarks, neither does it discuss hyper-parameter training nor the feature selection process.

ANN and SVR based forecasts are also compared in \cite{garulli2015mat} and in \cite{grolinger2016eff}. While Garulli et al.~\cite{garulli2015mat} find that SVM perform better than ANN, \cite{grolinger2016eff} finds the opposite. 

Tree-based approaches are also popular (see Section~\ref{sec:trees_and_forests}). Yunusov et al.~\cite{Yunusov2018sss} develop three two-day ahead forecasts based on a seasonal regression model, RF and SVR for nine feeders in the UK network and the forecasts are used in storage system control algorithms. A seasonal linear regression model outperforms the other two approaches in terms of forecast error and also the objective of the storage control. Nawaz et al.~\cite{nawaz2019aaf} uses Recursive Feature Elimination to select features and then compare various methods, i.e., kNN, RT, RF and SVR. They show that at different forecast horizons different methods perform best. In contrast, Cerquitelli et al.~\cite{Cerquitelli2019esm} compare ridge regression, ANN, and RF for hour-ahead building-level load forecasts and find that ANN generates the most accurate forecasts overall.

In their rather innovative work, Cordova et al.~\cite{Cordova2019cet} model residential load for the city of Tallahassee combining smart electricity meter data with transport network data and use information theory and causality models for the simultaneous study of the two datasets. They compare several statistical and machine learning algorithms, such as ARIMA, MLR, absolute shrinkage and selection operator (LASSO) and ridge regression, deep neural networks (DNN) and SVR.

In addition to ANN and SVR methods, Idowu et al.~\cite{idowu2016aml} also train, evaluate and compare MLP and Regression Tree for load forecasting in district \textit{heating} substations in Sweden. 
The authors show that SVM is the best performing method on their dataset with ANN and MLP following closely. Also, it is shown that the substations' internal state variables are important for high accuracy.
As shown,  linear regression models can also be quite accurate, but the choice of predictors is usually largely manual and ad-hoc. Thus, it would be beneficial to see studies with more automated feature selection, including LASSO methods which have been shown to be powerful at higher voltages (see e.g. \cite{Ziel14}).

Mirowski et al.~\cite{mirowski2014dfi} presents results from a uniquely large hierarchical data set of 32000 end-users over two years. They compare different approaches ranging from simple linear regression, to time series-based models (ARIMA and Holt-WinterS), a state-space model and machine learning (weighted kernal regression, kernel ridge regression and SVR) for forecasting of smart meter data at the customer, feeder, and system level. They find that SVR and the Holt-Winter model perform best in one-step ahead and day-ahead forecasting and for all aggregations. Simple combinations, like the mean of the best models, can improve the results (see also next Section).

\subsection{Hybrid, Combination and Ensemble Approaches}
\label{hybridSection}
The majority of the papers reviewed in this paper considers only individual models but combinations of models can produce much more accurate forecasts. The most popular and well-known methods for combining models are the ensemble approaches such as random forest, boosting and bagging, which combine weaker models to produce a strong predictor. Jung et al. \cite{Jung2020bem} propose a bagging ensemble of ANN for three building clusters in Seoul, South Korea, demonstrating that their methodology delivers better performance than other imputation techniques. In \cite{Laurinec2019due} bagging is applied for forecasting clusters of households which are then clustered with unsupervised ensemble learning methods to produce the final day-ahead aggregate load forecast.

Random forests (RF) are an ensemble learning method based on regression trees (Section \ref{sec:trees_and_forests}). In the regression case, RF constructs an ensemble of simple RT and make a final prediction based on the average of the individual trees. Compared to RT, they are typically more accurate but sacrifice interpretability. However, RF can handle a large number of features and can be used to analyse feature importance. For instance, Bogomolov et al.~\cite{bogomolov2016ecp} develop a novel RF regression model for predicting average and peak energy demand over the next seven days at the aggregate level, in Trentino (Italy), from energy data and aggregate mobile data usage. The authors extract several characteristics (e.g. mean, median, std, kurtosis, entropy) from the telecommunication data resulting in initially 3,000 features. After feature selection based on the total decrease in node impurities, the features space is reduced to 32 features with the number of customers per grid power-line having the highest feature importance. Kiguchi et al.~\cite{Kiguchi2019pil} use RFs to model residential load under time-of-use tariffs, considering different variables such as gender, age group, social class, and the number of other residents. They found RF to be more accurate than linear regression and ANNs and used RF to provide feature rankings based on the Gini coefficient.

Ben Taieb et al. \cite{taieb2016fui} consider an additive quantile regression model with boosting for probabilistic forecasts at an individual smart meter and aggregated level (all 3639 smart meters from the Irish CER dataset~\cite{Commission2012csm}). They compare this method to a standard additive model assuming normal distribution of errors. At an individual household level the proposed methods is most accurate, but at the aggregated the normal errors model is best. This is one of few demonstrations of the central limit theorem and shows this assumption of normality breaks down at the LV level. Nespoli et al. \cite{nespoli2020hdf} consider an XGBoost algorithm to perform hierarchical forecasting (just two levels LV to MV) and different ways to reconcile the forecasts. Ruiz-Abellón et al. \cite{ruiz-abellon2018lff} consider regression trees, and compare this with other ensemble approaches including bagging, random forest, conditional forest, and boosting, with random forest outperforming others, and XGBoost also showing promise. 

The ensemble approaches above use the same baseline models to create the final forecast. Another approach is to simply combine different models, with the aim to create the combined model that outperforms the individual ones. This is considered by Grmanová et al. \cite{grmanova2016iel} in which several machine learning and statistical baseline models, including SVR, MLR, AR, Holt-Winters, and a feed-forward ANN, are all averaged together and shown to perform statistically better than any individual model. An average of models is also considered by Dong et al. \cite{dong2016ahm}. The interesting feature of this paper is that five traditional machine learning models (ANN, SVR, least-squares SVM, Gaussian Process regression and Gaussian mixture model) are combined with a physics-driven model to forecast the air-conditioning (AC) component of the household electricity consumption. 

Not all papers use a simple average, Kodaira et al. \cite{kodaira2020oes} use a weighted average (learnt using a particle swarm optimisation) to combine a k-means and an ANN method. In \cite{tajer2017lfv} a learning-based load framework is proposed for an hour ahead forecast where the aggregated load prediction is combined from several weighted predictors for local sub-networks using kernels that are most suited to their load characteristics. Investigation of weightings in more detail would be an interesting avenue of further research. 

A slight modification to the simple averaging of models is to use a hybrid approach, where different features/components of the forecasts are generated using different models. These are by far the most common form of multiple-model based forecasts. Massidda and Marrocu~\cite{massidda2019smf} split the forecast into long term effects (trend and seasonality) and short term effects which are modelled by a random forest and multiple linear regression, respectively. In \cite{hashim2019anl}, the authors apply a hybrid neuro-fuzzy method with three stages to remove 
load shedding effects in load forecasting at the feeder level applied to two LV feeders from two substations in Najaf city (Iraq). % 
Amato et al. \cite{Amato2021fhr} propose a hybrid model based on a partially linear additive model to generate point forecasts for the Irish CER dataset~\cite{Commission2012csm}. They focused on forecasting for a small aggregation of households. 
The authors demonstrate that their modelling approach based on wavelets and spline decomposition 
can accommodate both the smooth  (temperature effects) and irregular (peak) patterns at the low aggregation levels. 

ARIMA and ANN type models are common components in model combinations. The authors in \cite{Bennett2014flv} propose a hybrid model using ARIMA and ANN to generate point forecasts for day-ahead total energy and peak load at the LV transformer level (serving 128 residential consumers). The authors report that ANN accounted more for small variations, while ARIMAX was more suited for modelling large spikes in load. This formed the justification for the authors to propose a hybrid model whereby load was first forecasted using ANNs, and if the forecasted demand was higher than a threshold, ARIMAX was used as a final forecasting model. Sulandari et al. \cite{sulandari2016fel} generate a hybrid model where the outputs from a TBATS (exponential smoothing state-space model with Box-Cox Transformation, ARMA error, trend and seasonal components) model are used as inputs to an ANN.  TBATS is used to decompose load time series data into a level, trend, seasonal and irregular components and ANN is applied to capture nonlinearity in the data. The hybrid model outperforms the individual TBATS and ANN models.   

Cao et al. combine the hybrid approach of ensemble techniques \cite{Cao2020hed}. They employ one bagging and five boosting techniques to improve the forecast accuracy of a deep belief network (DBN) and propose a hybrid modelling approach that adaptively combines six base ensemble technique, whereby the combination weights were computed via a k-nearest neighbour. The authors consider two sources of prediction uncertainty: (i) model misspecification, and (ii) data noise, both of which are assumed as being independent and as confirming to a Gaussian distribution. This could potentially be a rather restrictive assumption if this methodology were to be scaled. To satisfy the stationarity requirements of bagging and boosting, the data was differenced, and assumptions checked using the Augmented Dickey-Fuller test.

Ai et al.~\cite{ai2019hpd} propose a computational ensemble approach based on evolution strategies to select and configure candidate forecast models and combine their outputs. The forecast is used for very short-term forecasting and outperforms the persistence forecast and the predictions of the individual candidate models.

Sun et al.\cite{sun2015aea} propose a forecasting method at the feeder-level that consists of a two-step method that selects the best model based on the characteristics of the load. If the node is classified as regular, it is forecasted using a simple seasonal moving-average model. Irregular nodes are forecasted using a wavelet neural network approach.

\subsection{Probabilistic Forecasting} 
\label{section_prob} 
LV demand typically exhibits higher volatility and less seasonality, as compared to the high voltage, or system demand see \cite{Haben2019stl} and Fig.~\ref{lvplots}. For these reasons, probabilistic forecasts are often more appropriate than point forecasts due to their representation of the underlying uncertainty. A probabilistic forecast encapsulates the uncertainty associated with the forecasts, often in the form of a full probability distribution, which allows for more informed decision-making, as opposed to the case when either a point forecast or forecasts for a discrete set of predefined quantiles are considered. Probabilistic forecasts at the LV level can have a range of applications including, efficient operations and control of the distribution grid, anomaly detection, early warning system for peak consumption, designing time-of-use tariffs, energy trading, and online power management of the micro-grid, to name a few.

Uncertainty in forecasts can be conveyed in a variety of forms, they can be simple prediction intervals \cite{kodaira2020oes}, quantile estimates \cite{gerossier2018rda}, full continuous distributions \cite{pinto2017mpf} or ensembles, and can be univariate or multivariate. The best choice often depends on the purpose of the forecast. 

Evaluation of probabilistic forecasts requires the use of proper scoring functions, such as the continuous ranked probability score (CRPS) \cite{taieb2016fui} or the pinball loss score \cite{wang2019pil}. For probabilistic forecasts, such proper scoring functions reward a forecast if it is sharp (has a narrower distribution), subject to calibration (statistical consistency between forecast distribution and observations). Probabilistic scoring rules have also been used to estimate the model parameters, to try and ensure that both model estimation and evaluation take into account the uncertainty associated with load forecasts \cite{Arora2016fes}. There are also some less common methods for evaluating probabilistic forecasts. In addition to using a normalised quantile score, Gerossier et al. \cite{gerossier2018rda} also consider the count of observations between successive quantiles, in other words, similar to the validation via the probability integral transform. Yang et al. \cite{yang2020bdl} also consider the infrequently used Winkler score, which measures the sharpness and unconditional coverage of prediction intervals. 
\subsubsection{Prediction intervals}
Prediction intervals are the least descriptive form of probabilistic estimates, since they only provide the range of a particular distribution of points, e.g. from 5\% to 95\%, but are very easy to interpret and describe to non-experts. Chaouch and  Khardani \cite{chaouch2015rcq} uses kernel regression (with Nadaraya-Watson weights) to generate a full continuous density function forecast for peak demand of a smart meter, but only use it to generate a prediction interval whose accuracy is assessed using MAPE.  Kodaira et al.\cite{kodaira2020oes} forecasts prediction intervals by fitting various models to the residuals, including an assumed Gaussian model, a Chebyshev inequality-based method and a sample-based method.

\subsubsection{Quantile regression and similar methods}
One of the most common approaches to generate a probabilistic forecast is quantile regression. They are typically less computationally expensive than generating the full distribution but can be arbitrarily descriptive by choosing sufficiently large numbers of quantile cut points. Wang et al. \cite{wang2019pil} produce a quantile regression by fitting an LSTM to a pinball loss function, and compares it to other quantile methods, including a quantile regression neural network, a quantile gradient boosting regression tree and a benchmark that assumes Gaussian errors. Assuming Gaussian errors is a common simple model used since it is entirely defined by two parameters. Similarly, Elvers et al.~\cite{Elvers2019spl} propose a pinball loss guided CNN and show that it outperforms linear quantile regression, needing only to train one model, instead of one model per quantile and step in the horizon. The authors in \cite{yang2020bdl} produce a quantile forecast based on an LSTM model with a clustering-based pooling method to prevent over-fitting.  Zufferey et al.~\cite{zufferey2020psf} compare a pinball loss guided ANN with a probabilistic kNN for net power load forecasting. Within the probabilistic kNN, the forecasts of the k nearest neighbours are optimally weighted to minimize the pinball loss. A Bayesian neural network is also used to put a prior and update the parameters on the LSTM layers. The method is compared to, and improves upon, several benchmarks including quantile regression forests, and gradient boosting quantile regression. 

Ben Taieb et al. \cite{taieb2016fui} use an additive quantile regression using boosting, where base learners for each variable are added at each iteration of the algorithm. This is tested on both the 3639 individual smart meters from the Irish smart meter trial~\cite{Commission2012csm}, as well as their aggregation. Gerossier et al. \cite{gerossier2018rda} produce several simple quantile forecasts models including empirical quantiles conditional on the period of the week, and an additive model with temperature forecast and historical load as the main inputs. 

The paper by Bikcora et al. \cite{bikcora2018dfo} is slightly more unusual in that, in addition to consider a quantile model, it also considers an expectile regression which is rarely seen in the load forecasting literature. This is essentially a quantile regression using squared differences rather than absolute differences between the estimate and the observations\footnote{Expectiles are to the mean, as quantiles are to the median}. Yang and Hong, \cite{Yang2019del} propose a deep learning ensemble framework for probabilistic load forecasting. The authors use the Irish CER smart electricity meter data~\cite{Commission2012csm} and generate forecasts for one-hour and one-day ahead. The authors employ DNN for forecasting and use LASSO-based quantile combination strategy to refine the ensemble forecasts, whereby the pinball loss and Winkler score are used for model evaluation. As benchmarks, they use different ML approaches including quantile regression forest, quantile regression gradient boosting, and quantile LSTM. 

\subsubsection{Density estimation}

Less common than quantile estimates are full density estimates. These give the most information for the distribution but unless they assume Gaussian errors (as considered in e.g. \cite{wang2019pil}) they are typically much more expensive to estimate. The aforementioned paper by Bikcora et al. \cite{bikcora2018dfo} also considers a density estimate using an ARMA-GARCH model. This is one of the few examples of a traditional econometric model being used in load forecasting but has the advantage of requiring relatively few easy-to-train parameters, since it models both a mean and variance model.

Kernel density estimates are more common for density forecasts and are the summation of kernel functions with specific bandwidths which must be trained for each conditional variable \cite{pinto2017mpf}, \cite{Arora2016fes}. The drawback to these methods is that each new parameter is very expensive to train and therefore only a few conditional variables can be included. Arora et al. \cite{Arora2016fes} compare several conditional KDE models to generate household level density forecasts using the Irish CER dataset~\cite{Commission2012csm}. They condition consumption on time-of-day information and include decay parameters to reduce the influence of older data. Interestingly, they estimate model parameters by minimizing the in-sample continuous ranked probability score (CRPS) and evaluate the models using a range of performance scores to evaluate point, quantile, and probabilistic forecasts.

\subsubsection{Comparison of different methods}
Haben et al. \cite{Haben2019stl} present the first large-scale study of real LV feeder data by investigating the effect of temperature (real and predicted) on the accuracy of probabilistic load forecasts. They employed a range of forecasting models including quantile regression, GARCH-type models and conditional kernel density estimation among others. Interestingly, it was demonstrated that the power-law relationship between feeder size and forecast accuracy (Figure \ref{lvplots}) does not hold true for some feeders, such as those with large numbers of overnight storage heating, large amounts of lighting etc.

\subsubsection{Combined and hierarchical modelling}
Despite the importance of probabilistic forecasts for LV level demand, they are relatively few in number and are far outnumbered with point-based methods. Of the 221 papers reviewed, 44 included some probabilistic forecast element, and as seen in this section many of the models are relatively simple.  However, recent research is beginning to investigate some important aspects of probabilistic forecasts such as model combination and hierarchical modelling. Wang et al. \cite{Wang2019cpl} considers a constrained quantile regression averaging (CQRA) which is a weighted average across several quantile regression models with weights found by solving a linear programming problem that minimises the pinball loss function. They consider a quantile regression neural network, quantile regression random forests, and quantile regression gradient boosting and compare nine different forecast combination schemes, with the CQRA performing the best for reducing forecast errors. 

Another important topic is hierarchical forecasting which aims to ensure forecasts at the lowest level aggregate coherently with the higher level. This is becoming increasingly important with the development of smart connected grids. Although producing coherent point forecasts is relatively simple it is much more complicated for probabilistic forecasts. Ben Taieb et al. \cite{Taieb2020hpf} produce hierarchical probabilistic forecasts by first generating an empirical copula method for each series in the hierarchy and then using a hybrid model to revise them to be coherent. The copulas also ensure that inter-dependencies are preserved. It is worth noting here that although aggregate forecasts can be generated by simply summing up the corresponding forecasts at the lower level, it has been shown that this bottom-up approach produces poor results \cite{Hyndman2011ocf}. 
Hierarchical forecasting has also been a common topic in the Global Energy Forecasting Competitions (GEFCom). In particular, GEFCom2014 considered probabilistic load forecasts at two levels with the loads mostly at 4kV and 12kV \cite{Gefcom2014}.

\subsection{Other methods} 

This section considers some methods which don't fall into the traditional categorisation of forecasts described above. This includes the utilisation of new data sources, different frameworks and techniques. 

The authors in \cite{mcqueen2004mcs} focus on modelling the maximum demand on distribution networks using Monte Carlo simulations. Particularly, their goal is to create representative load profiles for use in the simulations. The daily and intraday demand is sampled from parametrized distributions. The method is applied to a single LV network and an entire network consisting of 557 transformers in Dunedin, New Zealand.

Zhou et al. \cite{zhou2019pse} use a graph approach where edges are correlation between variables and nodes are sectors (industrial and commercial firms in Shanghai). Nodes of the graph represent different sectors and their sizes correspond to the scale of the sectors, measured by average daily demand. A Granger causality test is used to indicate if there is a relationship between sectors and Akaike information criterion is used to choose lags. Pruning of the network is done using a minimum spanning tree to retain the most important information in the graph. Selected nodes are used as inputs to linear regression models.

Finally, the authors in \cite{xu2020ats} produce a day-ahead forecast of a twenty-floor office building in Shanghai using a gray theory model (numerically solved ordinary differential equations) using historic and current load. 
These methods demonstrate that there is a wide scope for applying less-traditional machine learning or time series methods for forecasting. These can only be expected to expand, especially as more granular or diverse datasets become available. 

\section{Trends in Forecasting}
\label{sec_trends_special_topics}

In this section, we discuss observed trends and several more specialised topics that we think are of interest to this research area, that may have received less attention, or are only just emerging.

\subsection{Process automation and move to operations}
An interesting emerging trend is attempts to generalise and automate a forecasting process, which is indifferent to the particular methods employed. A few papers, e.g. \cite{hoverstad2015stl,opera2019mla}, are investigating ways to produce an automated process including pre-processing and a feature selection process agnostic to the type of forecast method being applied. In \cite{hoverstad2015stl}, seasonalities are removed in a preprocessing phase, then an evolutionary algorithm is used to select features and train coefficients across a wide parameter space. In \cite{opera2019mla}, an automated process utilises three feature selection algorithms to pick inputs (LASSO, recursive feature estimation and univariate selection - the latter, considers F scores between input and response variables - the greater the F score, the greater the dependency of load on the input variable). After this step, clustering is applied and forecasts of each cluster are selected automatically from a set of various statistical and machine learning methods. In \cite{diamantoulakis2015bda}, a limited selection of pre-processing methods for short-term, smart grid, load forecasts are reviewed, considering only dimension reduction and clustering. Dimension reduction techniques include random projection of smart meter data, and online dimension reduction are applied, usually aiming to find correlations between features such as voltage, frequency, current, etc. The authors in \cite{tornai2016cfc} compare several forecasting methods, i.e., linear regression, radial basis function and ANNs, for classifying individual customers' time series. In particular, several classifiers are developed from an annotated dataset. A new unlabelled time series passes through the forecasts and it is assigned to the class with the lowest forecasting error.

With the increase and advances in monitoring, and communications equipment there are opportunities for more real-time applications, and online forecasting methods which are more robust to concept shifts. Of the reviewed papers only the online forecasting as seen in De Silva et al.~\cite{desilva2011ipc} utilising data-streams was found to consider such real-time operational questions. 

Research and comparison of different methods for data preprocessing, feature and parameter selection, and evaluation of different forecasting methods are of great importance and we hope to see more papers coming in this area in search of reliable, and robust optimal methods, that can be scaled up, productionised and used in real-life applications.

\subsection{'Divide and conquer' - Components of Load}
Splitting demand into different classes or clusters is a common technique to improve the aggregated forecasts that manifest in many different approaches. We discuss clustering in more details in Section \ref{sec:clustering}, but here we list several examples of splitting load into different components to improve the accuracy of the prediction. There are many ways to split up LV level demand, such as according to particular appliances, individual consumers, or the time series itself can be split into its frequency components as is common in Wavelet type models (some of these types of models are described in Section \ref{secStatandTS}). 

In \cite{dong2016ahm} demand is split into air-conditioning (AC) demand and non-AC demand and a hybrid model is used to predict a total load combining different data-driven machine learning algorithms and physics-driven heating models. 

Net load forecasting \cite{chu2017nlf} presents an interesting adaptation to traditional load forecasting. Preprocessing includes decomposition of the net-load time series to remove low-frequency load variation due to daily human activities. The load is split into daytime and nighttime and models are trained separately on each component. The exogenous predictors for the daytime forecast's include sky image features.  

Rabie et al. \cite{rabie2019afb} introduces a feature selection methodology 
for peak load forecasting at the grid level. The interesting aspect is that the forecasting is performed as a classification task, not a regression, classifying load into three classes; low, medium and high. The features selection task is a two-step method consisting of a filter feature selection which ranks the features based on several criteria and a feature wrapper method on the ranked features of the previous step using a Naive Bayes classification. However, the proposed feature selection technique is applied to a small set of features (five), selecting three of them. The method should have been tested in a larger feature set.
Similarly, in \cite{vats2020meo}, the authors perform a day-ahead load forecast for buildings in Patna, India, using classification. Particularly, they classify the loads into five classes and use them to predict the load for the next day using regression analysis. Three classification methods were considered, kNN, Random Forest and SVM, using the power factor, voltage, current, weekday and hour features.

For a special case of load splitting into appliances (so-called `behind the meter') an earlier analysis by \cite{bao2011ubp} based on only one household with 60 days of data, aimed to forecast user behaviour, i.e., devices usage. Regular devices (such as fridges) are more accurately forecasted using simple periodic models, while for other devices a semi-Markov model performed better. Hence, they find that the overall best predictions are done through a hybrid model. Welikala et al.~\cite{welikala2017iau} propose an approach using appliance usage patterns to improve the performance of non-intrusive load monitoring. They also show, how their approach can be used for conducting more accurate very-short term load forecasts at the household-level (5 minutes ahead).

\subsection{Peak forecasting}  \label{sec:application_specific}

Peak forecasting is relevant for demand forecasts for several reasons: 1) errors when forecasting peaks are usually much more costly than other errors; 2) peaks (in 'values over the threshold' sense)  are relatively rare compared to the full time series,  and 3)  it is more difficult to predict peaks in LV settings than in more aggregated, smoother demand. However, only a few papers discuss peaks forecasting and the related errors,  (see e.g. \cite{chaouch2015rcq, Jacob2020faa}). In \cite{Komatsu2020pda} early warning systems for peak electricity consumption demand is presented using SVR for demand forecasting, where demand alerts are sent to grid managers when the predicted demand exceeds the predefined threshold.

\subsection{Forecast Evaluation}
\label{subsec:eval} 
Issues around peak significance spill over into forecast evaluation. Over the last few years, numerous methodologies have been proposed to forecast the household-level consumption time series. However, despite the significant advancements in the modelling of smart electricity meter data, very little progress has been made in designing reliable performance scores to evaluate high-resolution household-level point forecasts. In \cite{Haben2014ane}, a new, adjusted error measure based on the $l^p$ norm, ($p$ is 4, to highlight peak errors) is proposed that evaluates the accuracy of a model for the timing and amplitude of the peak at the household-level. They show that a flat forecast can outperform a better informed, `peaky' forecast, if evaluated using the absolute error. Namely,  a forecast that accurately predicts a peak's amplitude and duration, but slightly displaced in time, e.g half-hour early, will incur a double penalty, one penalty for predicting a slightly early non-existing peak and another for missing the peak one half-hour later. Teeraratkul et al.~\cite{teeraratkul2017sba} show that Dynamic Time Warping (DTW), a popular time series measure can also be used as a forecast error measure and propose a shape-based forecasting approach to minimise it. While the measures proposed by these authors \cite{Haben2014ane, teeraratkul2017sba} is a step in the right direction, there is a need for similar error measures that are rewarding more informed forecasts that accurately predict peaks at the LV level. Any new error measures can be used to generate tailored forecasts as with the Average Adjusted Forecaster (AAF) proposed in \cite{Haben2014ane} or simple kNN approaches as in \cite{voss2018adjusted} that are aimed to minimise the adjusted $p$-norm error. Rowe et al~\cite{rowe2014apr} show how the AAF can be used in a peak reduction algorithm for battery control in LV networks. Voss~\cite{voss2020pbr} showed how optimal choice and configurations of the error measure, depend on the specific down-stream optimization objective for household-level energy management. More studies are needed in these specific downstream applications.

Finally, it is encouraging to see that statistical tests are becoming a more common trend in forecast evaluation, e.g. \cite{grmanova2016iel} uses the Wilcoxon rank sum test. In \cite{lusis2017str}  a Holm-Bonferroni's multiple hypothesis test based on a one-sided Wilcoxon signed-rank test is used to see if using calendar effects significantly reduces the error (it does not). Friedman statistical test and nonparametric Wilcoxon test are applied in \cite{gerossier2018rda} to see if the models perform differently (they do), however the best performing model is not statistically different in performance to one of the reference models.

\subsection{Clustering for Forecasting Improvement}
\label{sec:clustering}

Clustering can be used to improve load forecasts in several ways. For instance, it can be used to estimate parameters of parametric models by fitting one set of parameters that suits similar households ~\cite{Arora2016fes}. Similarly, deep learning models can benefit from additional training data to pool profile uncertainties, allowing more data of similar households to train (possibly different) models~\cite{Shi2017dlf,yang2020bdl} (see also Section~\ref{sec:deep_learning}). Further, load profile clustering can be used to improve an aggregate forecast. This is done by first agglomerating similar groups, generating a forecast per group and then aggregating the groups'  forecasts~\cite{Wijaya2015caf,Humeau2013elf,kurniawan2015cba,fahiman2017ilf,gajowniczek2018sso,alzate2013iel}. The two main advantages of this approach are that \begin{itemize}
	\item
	the clusters of similar behaviour result in more regular time series, facilitating accurate prediction;
	\item the load becomes smoother, so even random clusters might bring forecast accuracy improvements (as noticed in \cite{Wijaya2015caf}).
\end{itemize}
The main challenge with clustering evaluation is that one needs to have sufficient disaggregated data to test the effect. As illustrated in Section~\ref{secdatasets}, there are a limited number of openly available smart meter dataset (beyond the Irish CER dataset~\cite{Commission2012csm} and UK Low Carbon London~\cite{UK2014ulc}) that can be used as benchmarks. Other datasets used in most papers are not open, so it is difficult to compare different methods. 

Load profiles are most commonly clustered directly using the raw time series or indirectly based on features obtained from the time series. When clustering load profiles, especially at the LV level, the choice of a distance measure can significantly influence the clustering result. 

Gajowniczek et al.~\cite{gajowniczek2018sso} compare different distance measures and find that the edit distance, longest common sub-sequence (LCSS), as well as the cross-correlation and TQuest distance\cite{TQuest} work well for clustering. Furthermore, they find that around 6-8 clusters can most improve the forecast results of ANN and SVR. 

We observed that broadly,  k-means is the most popular method. More recently, methods better suited to clustering time series data are emerging such as k-shape. The methods are typically used as a pre-processing step to improve short term forecasts. 

\subsubsection{K-means clustering}
Humeau et al.~\cite{Humeau2013elf} analyse if clustering can improve the forecast of the aggregated Irish CER dataset~\cite{Commission2012csm} using \textit{k}-means. They find that for MLP and linear regression clustering does not improve results, but the SVR results are best when using four clusters. In \cite{quilumba2015usm}, the authors also use k-means clustering of smart-meter data to produce aggregate forecasts. Once clusters are found, ANNs are used to forecast time-horizons up to one-day ahead. In \cite{Shahzadeh2015ilf}, neural networks are used to forecast the Irish CER dataset~\cite{Commission2012csm}. They show that clustering consumers result in more accurate forecasts. They train a separate neural network model on each separate cluster, whereby for clustering, they use \textit{k}-means.

The authors of \cite{Li2016slf} model the Irish CER smart meter data and argue that their modelling takes into account different consumer behaviours. Firstly, average load profiles based on different day types are calculated. Secondly, consumers with similar consumption behaviour are clustered using \textit{k}-means. Finally, an Online Sequential Extreme Learning Machine (OS-ELM) is used for load forecasting for different clusters, which are aggregated to get the system load. 

In \cite{li2020sbp}, the authors use a parallel \textit{k}-means implementation in Spark for clustering users. Then they forecast the clustered demand using essentially a parallel implementation of an ANN with a hidden layer of 20 neurons. The clustering and parallelisation attempt to address the increasing volume of household data. The method is applied to the Irish household dataset. Lu et al. in  \cite{lu2019awl} propose a Davies-Bouldin index-based adaptive \textit{k}-means algorithm to cluster 2000 large users, in Foshan, Guangdong province of China, into several groups. A hidden Markov model describing the probabilistic transitions of different load levels is established for each cluster to extract the representative dynamic weekly load features. An MLR and ANN are used as benchmarks, however, the HMM does not outperform the benchmarks for all clusters. The authors suggest that a combination of forecasts should be utilized to improve forecast accuracy.

While a vast majority of studies on modelling residential electricity data use only lagged consumption values, \cite{Fu2018csl} include temperature along with consumption, and report that temperature was a salient feature in the modelling. Their modelling relies on clustering consumers using a Fuzzy C-means clustering algorithm. 

Different clusters are modelled separately, and the individual predictions for a given cluster are aggregated to compute the net usage. 
Abera and Khedkar \cite{abera2020mla} focus on forecasting the appliance consumption and peak demand, using the Irish CER~\cite{Commission2012csm} and UMass datasets. They first cluster the profiles using CLARA (Clustering LARge Application - an extension of k-medoids), while for forecasting they use SVMs and ANNs.

\subsubsection{Other methods}
The work \cite{fahiman2017ilf} focuses on the short term load forecasting of the aggregate households one day-ahead. The authors first cluster the household profiles of the Irish CER dataset~\cite{Commission2012csm} using k-shape clustering, then they forecast the load of the cluster before applying a \textit{weighted} sum of the clusters to forecast the aggregate household demand. For forecasting two main methods were used, ANN and Deep Belief Networks (DBN), a multi-layer generative model that learns one layer of features at a time from unlabelled data.
For clustering, a \textit{k}-shape algorithm is used, which clusters households with similar shape, instead of the traditional \textit{k}-means. 
The \textit{k}-shape with DBN outperforms all other methods (\textit{k}-means and/or ANN).

The authors of \cite{Fu2018csl} propose a short-term residential load forecasting framework. In their modelling, they consider the adoption of increasing-block tariffs, which is broadly based on the concept of allocating each consumer to a consumption block such that higher consumption blocks are associated with higher prices. 

To forecast the system load for a group of customers, \cite{Goehry2020ame} employ clustering techniques, aggregation methods, and machine learning along with survey and weather data to produce a forecast applied to the Irish CER dataset~\cite{Commission2012csm}. For clustering consumers, the authors use and compare hierarchical agglomerative clustering (a flexible technique, giving different numbers of clusters on different hierarchical levels) and random clustering. A random forest is then used for forecasting. 
Interestingly, the authors reported that random clustering performed as well as hierarchical agglomerative clustering, which may be due to the rather similar nature of consumers considered. Namely, hierarchical agglomerative clustering may be more suited for applications where there is more heterogeneity in the predictor variables being considered. Chen et al.\cite{chen2019daa} propose a multi-step load forecasting approach for aggregate load in the LV grid. They propose to use Affinity Propagation clustering to separate the customers into similar groups of customers (see section \ref{sec:deep_learning}).

Kernel spectral clustering is used in \cite{alzate2013iel}, to improve aggregated forecasts (Periodic auto-regressive base forecast) of Irish smart meters~\cite{Commission2012csm}. Small numbers of clusters appear to perform best. In \cite{kurniawan2015cba} the authors cluster the Irish CER data using features with the highest correlation.  Several clustering strategies are deployed, including a random assignment. As in \cite{alzate2013iel}, the accuracy changes with the number of clusters, but this paper also shows that it can depend on the forecast and cluster method as well. There does not appear to be any criteria for knowing this in advance.

\section{Low-Voltage Load Forecasting Datasets} 
\label{secdatasets}

A number of interesting features were discovered about the data in the reviewed 221 papers. Firstly, only 52 use at least one openly available datasets to illustrate the results, i.e. less than 24\% of the journals presented results that could be potentially replicated by the wider research community. Of these 52 papers using open data, 22 (or $42\%$)  of them used the Irish CER Smart Metering Project data~\cite{Commission2012csm}, four used data from UK Low Carbon London project~\cite{UK2014ulc}, four from Ausgrid\footnote{\url{https://www.ausgrid.com.au/Industry/Our-Research/Data-to-share/Solar-home-electricity-data}} and three used the UMass dataset. In other words, out of the papers using open data, $56\%$, presented results that used data from only four open data sets. 

The overuse of a particular dataset can result in biases (both conscious and unconscious) where methods are developed and tested but the features of the data may be well known or familiar from overuse. In these cases a scientifically rigorous experiment is impossible. Further, reliance on a single dataset (especially those which are no more than 2 years in length like the Irish CER dataset~\cite{Commission2012csm}) risks the development of models which may be based on spurious features and patterns and may not be representative of the wider energy system. This can be alleviated somewhat by including multiple open data sets, as it was done in some of the papers reviewed (\cite{abera2020mla, Laurinec2019due, Wang2018aef}). To support the LV forecast research community the authors are going to share a modifiable list of open data sets at the LV level. A list and some of the properties of the data are shown in Table \ref{tab:datasets}.  

With the rapid change in low carbon technologies being connected to the grid and home, new energy efficiency interventions, and adjustments in demand usage behaviours, demand data can quickly become irrelevant or unrepresentative. Further, such data sets are based on trials where participants are subject to incentives or other interventions. For example, different tariffs were considered for some households in the Irish CER dataset~\cite{Commission2012csm}. This means that their demand may not represent `normal', every-day behaviour.  

There are now some initiatives that are attempting to solve some of the issues of sparse and intermittent open data produced by limited innovation projects. An example of this is the Smart Energy Research Lab from UCL in the  UK\footnote{\url{https://www.ucl.ac.uk/bartlett/energy/research/energy-and-buildings/smart-energy-research-group-serg}} which is attempting to make smart meter data (as well as other useful data sets associated with the same homes) available on an ongoing basis for research from a wider range of participants as well as a large control group of households which will not participate in any initiatives or trials. 

\subsection{Data resolution}
Resolution is another important aspect of the data. Half-hourly data is the standard resolution of smart meter data although it can be as low as 10 or 15 minutes. The smallest stated resolution of the data within the papers had 48 with hourly data, 51 half-hourly, 23 as 15 minutes and 8 as 10 minutes. There was also a large number of papers where the resolution was not clear or no real data was presented (62 papers). Data with resolutions of between ten minutes and an hour are probably sufficient for demand control applications and are likely representative of what data is available in practice. However, this isn't sufficient for more high-resolution applications such as voltage control. In fact, only 11 papers considered data of resolution of 1 minute or less. This could pose a difficulty for validating the common voltage and Var control application in this review (see Section~\ref{sec:LVLF-applications}). A large number of papers where the resolution is not clear should also be a concern, as this prevents recreation of the results. 

\subsection{Forecast horizon}
Another crucial aspect of this review is the forecast horizon. Different horizons are useful for different applications. Short term (day to the week ahead) are typical for operational time scales, whereas long-term forecasts (over a year) are more useful for planning. The majority of the papers reviewed were at a short term time scale with 80 of the papers considering day ahead forecasts, another common horizon was an hour ahead. Very few papers went at shorter horizons than an hour (twelve). There were slightly more papers that forecast beyond a day (16 were between 2 days and a week ahead) and only 13 papers were at horizons of a month or more. Once again there was a large number of papers, 80 in total, where the horizon length was not identifiable. 

\subsection{Overview of LV datasets}
As we have already mentioned, the current choice of open datasets that can be used for a benchmark is very limited, relying mostly on the CER Irish smart meter data, which is now a decade old and has some selection bias limitations (most of the houses have 3-4 bedrooms etc.). In order to continually expand research in this area, a strategy is required to regularly open more diverse datasets, converging towards common formats and standards and clarifying licences and terms of usage. Clear license information is especially relevant for industry-based research. We have established a list of open datasets (see Table~\ref{tab:datasets} and \url{https://low-voltage-loadforecasting.github.io/}) with the hope that it will continue to grow, and that new methods for privacy and safety protection (anonymisation, aggregation, synthetic 'look alike' datasets etc.) will enable more availability of datasets in the future. Finally it is vital to provide proper and thorough documentation with the data sets. The quality of the datasets is not clear and in many cases any preprocessing or data-cleaning techniques are not provided with the data.

Most datasets are at the residential level collected from smart meters. More diverse datasets of non-residential customers and different grid levels (substations and transformers) are needed for better LV forecasting research. 

Some datasets that have been cited in the literature like PLAID~\cite{Medico2020ava} OCTES, BLUED~\cite{Anderson2012baf},  and DRED~\cite{UttamaNambi2015lle} were offline at the time of writing. The Pecanstreet Dataport~\cite{Pecan2018d} database was once publicly available for research but then closed, and now only a subset is still accessible. Other datasets are available for download but are hard to trace, as identifiers like a DOI or a paper to cite are not available. All of these obstruct the reproducibility of the research. Therefore, new datasets should be published on archiving platforms like IEEE data port~\footnote{\url{https://ieee-dataport.org/}}, Zenodo~\footnote{\url{https://zenodo.org/}}, Figshare~\footnote{\url{https://figshare.com/}}, or  arXiv~\footnote{\url{https://arxiv.org/help/submit\#datasets}}. A recent contribution to more reproducible time series research is the Monash Time Series Forecasting Archive~\cite{godahewa2021mts}. It contains the dataset of the UK Low Carbon London trial~\cite{UK2014ulc} and the UCI datasets~\cite{candanedo2017ddp, Hebrail2012ihe}.

\begin{sidewaystable*}
	\tiny
	\caption{Overview of Low-voltage Load Datasets (see online version for embedded hyperlink to the data set by clicking on the name).} \label{tab:datasets}
	\resizebox{!}{0.9\height}{
		\begin{tabular}{p{0.16\linewidth}p{0.05\linewidth}p{0.04\linewidth}p{0.05\linewidth}p{0.04\linewidth}p{0.04\linewidth}p{0.02\linewidth}p{0.02\linewidth}p{0.12\linewidth}p{0.14\linewidth}p{0.14\linewidth}}
			\\
			\toprule
			Name & Type  & No. Customers & Resolution & Duration & Intervention & Sub-metering & Weather avail. & Location & Other data provided & Access/Licence \\
			\midrule 
			\href{https://www.ea.tuwien.ac.at//projects/adres_concept/EN/}{ADRES}~\cite{einfalt2011kfa}
			& Households & 30    & 1 s   & 2 weeks & None  & No    & No    & Austria (Upper Austria) & Voltage & Free for Research (E-Mail) \\
			\href{https://www.ausgrid.com.au/Industry/Our-Research/Data-to-share/Solar-home-electricity-data}{Ausgrid Solar Home} &  Households &  300 & 30 min &   3 years & None &  No &  No &  Australia (NSW) &  &   No Licence \\
			\href{https://www.ausgrid.com.au/Industry/Our-Research/Data-to-share/Distribution-zone-substation-data}{Ausgrid substation data} &   Substation & 225 & 15 min &  20 years & None &  No &  No &  Australia (NSW) &  & No Licence \\
			\href{https://sourceforge.net/projects/greend/}{GREEND Electrical Energy Dataset (GREEND)}~\cite{monacchi2014gae} & Households & 8     & 1 s   & 3-6 months & None  & Yes   & No    & Austria, Italy & Occupancy, Building type & Free (Access Form) \\
			\href{https://archive.ics.uci.edu/ml/datasets/Appliances+energy+prediction}{UCI Appliances}~\cite{candanedo2017ddp} & Households & 1     & 10 min & 4.5 months & None  & No    & Yes   & Belgium (Mons) & Lights, Building information & Free (No Licence) \\
			\href{https://ieee-dataport.org/open-access/industrial-machines-dataset-electrical-load-disaggregation}{INDUSTRIAL MACHINES}\cite{Bandeira2018imd}
			& Industrial & 1 & 1 Hz  & 1 month & None  & Yes   & No    & Brasil (Minas Gerais) &       & CC BY \\
			\href{https://dataverse.harvard.edu/dataset.xhtml?persistentId=doi:10.7910/DVN/ZJW4LC}{Rainforest Automation Energy} \cite{Makonin2018rtr} & Households & 2     & 1 Hz  & 2 months & None  & Yes   & Yes   & Canada & Environmental, Heat Pump,  & CC BY \\
			\href{https://dataverse.harvard.edu/dataset.xhtml?persistentId=doi\%3A10.7910/DVN/FIE0S4\%20}{AMPds2} \cite{Makonin2016ata, Makonin2016ewa}  & Households & 1     & 1min  & 2 years & None  & Yes   & Yes   & Canada (Alberta) & Gas, Water, Building Type and Plan & CC BY \\
			\href{https://carleton.ca/sbes/publications/electric-demand-profiles-downloadable/}{Sustainable Building Energy Systems 2017} \cite{Johnson2017edf} & Households & 23    & 1 min & 1 year & None  & Yes   & No    & Canada (Ottawa) & Sociodemographic (Occupants, Age, Size) & Free (Attribution, E-Mail) \\
			\href{https://carleton.ca/sbes/publications/electric-demand-profiles-downloadable/}{Sustainable Building Energy Systems 2013} \cite{Saldanha2012mee} & Households & 12    & 1 min & 1 year & None  & Yes   & No    & Canada (Ottawa) & Sociodemographic (Occupants, Age, Size) & Free (Attribution, E-Mail) \\
			\href{https://data.lab.fiware.org/organization/9569f9bd-42bd-414f-b8d9-112553ea9dfb?tags=FINESCE}{FINESCE Horsens}
			& Households & 20    & 1 h   & several days & None  & Yes   & Yes   & Denmark (Horsens) & EV, PV, Heat Pump, Heating, Smart Home,  & CC BY-SA \\
			
			\href{https://archive.ics.uci.edu/ml/datasets/Individual+household+electric+power+consumption}{UCI Individual household electric power cons.}~\cite{Hebrail2012ihe} & Households & 1     & 1min  & 4 years & None  & Yes   & No    & France (Sceaux) & Reactive Power, Voltage & CC BY \\
			\href{https://mediatum.ub.tum.de/1375836}{BLOND-50}~\cite{Kriechbaumer2017bbo} & Commerical & 1     & 50 kHz & 213 days & None  & Yes   & No    & Germany &       & CC BY \\
			\href{https://mediatum.ub.tum.de/1375836}{BLOND-250}~\cite{Kriechbaumer2017bbo} & Commerical & 1     & 250 kHz & 50 days & None  & Yes   & No    & Germany &       & CC BY \\
			\href{https://zenodo.org/record/3855575\#.YKQgGKgzaUk}{Fresh Energy}~\cite{Beyertt2020fzb} & Households & 200   & 15 min & 1 year & Behaviorial & Yes   & No    & Germany & Agegroup, Gender of main customer & CC BY \\
			\href{https://data.lab.fiware.org/organization/9569f9bd-42bd-414f-b8d9-112553ea9dfb?tags=FINESCE}{FINESCE Factory} 
			& Industrial & 1     & 1 min & 2 days & None  & Yes   & No    & Germany (Aachen) & Machines & CC BY-SA \\
			\href{https://pvspeicher.htw-berlin.de/wp-content/uploads/MFH-Lastprofil_2014_17274_kWh.csv}{HTW Lichte Weiten}~\cite{htw2019ldb}
			& Households & 1 building & 15 minute & 1 year & None  & No    & No    & Germany (Berlin) &       & Free (No Licence) \\
			\href{https://pvspeicher.htw-berlin.de/veroeffentlichungen/daten/lastprofile/}{HTW Synthetic}~\cite{Tjaden2015rel} & Households & 74    & 1 s   & 1 year & None  & No    & No    & Germany (Representative) & Synthetic dataset merging & CC BY-NC \\
			\href{https://data.open-power-system-data.org/household_data/}{CoSSMic} \cite{Open2020dph} & Households, SME & 11    & 1min, 15min, 1H & 1-3 years & None  & Yes   & No    & Germany (South) & PV, EV, Type (Residential/SME) & CC BY \\
			\href{https://im.iism.kit.edu/sciber.php}{SciBER}~\cite{Staudt2018san} & Municipal & 107   & 15min & 3 years & None  & No    & No    & Germany (South) & Type (Office, Gym, ...) & CC BY \\
			\href{https://iawe.github.io/}{iAWE}~\cite{batra2013idi}
			& Households & 1     & 1 Hz  & 2 months & None  & Yes   & No    & India (New Delhi) & Water & Free (No Licence) \\
			\href{https://combed.github.io/}{COMBED}~\cite{Batra2014aco} & Commerical & 1     & 30 s  & 1 month & None  & Yes   & No    & India (New Delhi) &       & Free (No Licence) \\
			\href{http://www.ucd.ie/issda/data/commissionforenergyregulationcer/}{Irish CER Smart Metering Project data}~\cite{Commission2012csm} & Households, SME, Other & 3835  & 30min & 1.5 years & Tariff & No    & No    & Ireland & Type (Residential/SME/Other) & Free (Signed Access Form) \\
			
			\href{https://github.com/Nikasa1889/ShortTermLoadForecasting}{Hvaler Substation Level data}~\cite{DangHa2017lst} & Substation & 20    & 1 h   & 2 years & None  & No    & No    & Norway (Hvaler) &       & Free (No Licence) \\
			\href{http://web.lums.edu.pk/~eig/CXyzsMgyXGpW1sBo}{Energy Informatics Group Pakistan}~\cite{Pereira2014sap} & Households & 42    & 1 min & 1 year & None  & Yes   & No    & Pakistan & Sociodemographic (building properties, no of people, devices) & Free (No Licence) \\
			\href{https://archive.ics.uci.edu/ml/datasets/ElectricityLoadDiagrams20112014}{UCI Electricity Load Diagrams}~\cite{Godahewa2021ehd}
			& Different & 370   & 15 min & 2 years & None  & No    & No    & Portugal &       & Free (No Licence) \\
			
			\href{http://www.vs.inf.ethz.ch/res/show.html?what=eco-data}{Electricity Consumption and Occupancy (ECO)}~\cite{Christian2014ted, Wilhelm2015hom} & Households & 6     & 1 Hz  & 8 months & None  & Yes   & No    & Switzerland & Occupancy & CC BY \\
			\href{https://www.gov.uk/government/publications/household-electricity-survey--2}{Household Electricity Survey (HES)}~\cite{Zimmermann2012hes} & Households & ~{}250 & 2 min & 1 month (255) to 1 year (26) & None  & Yes   & No    & UK    & Consumer Archetype & Request \\
			\href{https://beta.ukdataservice.ac.uk/datacatalogue/studies/study?id=8634}{METER}~\cite{Grunewald2019muh} & Households & 529   & 1 min & 28 hours & None  & No    & No    & UK    & Activity data, Sociodemographic & Free for Research (Access Form) \\
			\href{https://datashare.ed.ac.uk/handle/10283/3647}{IDEAL Household Energy Dataset}~\cite{goddard2020ihe}
			& Households & 255   & 1 s     & 3 years & None  & Yes   & No    & UK    & Smart Home, Sociodemographic, energy awareness survey, room temperature and humidity, building characteristics & CC BY \\
			\href{http://www.networkrevolution.co.uk/resources/project-data/}{Customer-Led Network Revolution project data}~\cite{sidebotham2015cln} & Households, SMEs & ~{}12000 & 30 min & > 1 year & Time of Use & No    & No    & UK    & EV, PV, Heatpump, Tariff,  & CC BY-SA \\
			\href{https://jack-kelly.com/data/}{UK Domestic Appliance-Level Electricity (UK-DALE)}~\cite{Jack2015tud} & Households & 5     & 16 kHz, 1s & months, one house > 4 years & None  & Yes   & No    & UK (London area) &       & CC BY \\
			\href{https://data.london.gov.uk/dataset/smartmeter-energy-use-data-in-london-households}{UK Low Carbon London}~\cite{UK2014ulc,Godahewa2021lsm} & Households & 5567  & 30min & 2 years & Time of Use & No    &  No   & UK (London) & CACI Acorn group & Free (No Licence) \\
			\href{https://www.refitsmarthomes.org/datasets/}{REFIT}~\cite{Murray2016rel, Murray2017ael} & Households & 20    & 8 s   & 2 years & None  & Yes   & Yes   & UK (Loughborough) & PV, Gas, Water, Sociodemographic (Occupancy, Dwelling Age, Dwelling Type, No. Bedrooms) & CC BY \\
			\href{https://www.spenergynetworks.co.uk/pages/flexible_network_data_share.aspx}{Flexible Networks for a Low Carbon Future} & Substations & Several Secondary  & 30 min & 1 year & None  & No    & No    & UK (St Andrews, Whitchurch, Ruabon) &       & Free (Access Form) \\
			\href{https://ukerc.rl.ac.uk/DC/cgi-bin/edc_search.pl?GoButton=Detail\&WantComp=146\&\&RELATED=1}{NTVV Substations} & Substation & 316   & 5 s   & > 4 years & None  & No    & No    & UK (Thames Valley) &       & Open Access (Any purpose) \\
			\href{https://ukerc.rl.ac.uk/DC/cgi-bin/edc_search.pl?GoButton=Detail\&WantComp=147\&\&RELATED=1}{NTVV Smart Meter} & Buildings & 316   & 30 min & > 4 years & None  & No    & No    & UK (Thames Valley) &       & Open Access (Any purpose) \\
			
			\href{https://site.ieee.org/pes-iss/data-sets/}{IEEE PES Open Data Sets} 
			& Households, Commercial & 15    & 1 min, 5 min, 15 min & 2 weeks & None  & No    & No    & USA   & Connection limit & Free (No Licence) \\
			\href{http://redd.csail.mit.edu/}{Reference Energy Disaggregation Data Set (REDD)}~\cite{Kolter2011rap} & Households &  ~{}10 & 1 kHz & 3-19 days & None  & Yes   & No    & USA (Boston) & Voltage & Free (Attribution, E-Mail) \\
			\href{http://wzy.ece.iastate.edu/Testsystem.html}{Iowa Distribution Test Systems}~\cite{Bu2019atd} & Substation & 240 nodes & 1 H   & 1 year & None  & Yes   & No    & USA (Iowa) & Grid data & Free (Attribution) \\
			\href{https://www.pecanstreet.org/dataport/}{Pecanstreet Dataport (Academic)}~\cite{Pecan2018d} & Households & 30    & 1min, 15min, 1H & 2-3 years & None  & Yes   & Yes   & USA (mostly Austin and Boulder) & PV, EV, Water, Gas, Sociodemographic & Free for Research (Access Form) \\
			
			\href{https://neea.org/resources/rbsa-ii-combined-database}{Residential Building Stock Assessment}~\cite{Larson2014jua} & Households & 101   & 15 min & 27 months & None  & Yes   & No    & USA (North West Region) & Building Type (Single Family, Manufactured, Multifamily) & Free (Access Form) \\
			
			\href{http://lass.cs.umass.edu/projects/smart/}{SMART* Home 2017}~\cite{Barker2012sao} & Households & 7     & 1 Hz  & > 2 years & None  & Yes   & Yes   & USA (Western Massachussets) &       & Free (No Licence) \\
			\href{http://lass.cs.umass.edu/projects/smart/}{SMART* Apartment}~\cite{Barker2012sao} & Households & 114   & 1 min & 2 years & None  & No    & Yes   & USA (Western Massachussets) &       & Free (No Licence) \\
			\href{http://lass.cs.umass.edu/projects/smart/}{SMART* Occupancy}~\cite{Barker2012sao} & Households & 2     & 1 min & 3 weeks & None  & No    & No    & USA (Western Massachussets) & Occupancy & Free (No Licence) \\
			\href{http://lass.cs.umass.edu/projects/smart/}{SMART* Microgrid}~\cite{Barker2012sao} & Households & 443   & 1 min & 1 day & None  & No    & No    & USA (Western Massachussets) &       & Free (No Licence) \\
			\href{http://lass.cs.umass.edu/projects/smart/}{SMART* Home 2013}~\cite{Barker2012sao} & Households & 3     & 1 Hz  & 3 months & None  & Yes   & No    & USA (Western Massachussets) & Solar, Wind, Environmental, Smart Home, Voltage,  & Free (No Licence) \\
			\bottomrule
		\end{tabular}
	}
\end{sidewaystable*}

\section{Low-Voltage Load Forecasting Applications}
\label{sec:LVLF-applications}

Forecasts are often used within specific applications, which require estimates for planning,  operation and trading. This section overviews some of the common applications encountered as part of the review. 

\subsection{Network Design and Planning}

Forecasts are often used in grid design optimisation. Dupka et al.~\cite{Dupka2011FSP} use a Gaussian distribution forecast to minimise the costs of locating and sizing capacitors on the distribution system. Ravadanegh et al.~\cite{ravadanegh2013hao} use medium and long term load forecasts to locate the optimal site and size of a distribution substation. The authors in \cite{kavousifard2014mop} use a Gaussian model to forecast the load uncertainty for distribution feeders with wind turbines to optimise the selection of the topology and position of sectionalising switches. Kavousi-Fard and Niknam \cite{Kavousi2014msd} use the forecast of active and reactive loads to improve the reliability of the distribution network by focusing on the reconfiguration of a distribution feeder. Ahmadigorji et al. \cite{ahmadigorji2009odp} study the optimisation of the location points of portable distribution generation based on a cost/worth analysis. Load forecast uncertainty is incorporated by assuming Gaussian distribution for mean and predicted annual peak load.

The work of \cite{Lee2020ncr} aims to reduce the neutral current via phase arrangement. An LSTM model was adopted for monthly load forecasting, and phase arrangement optimisation was performed using particle swarm optimisation.

\subsection{Network Operations and Control}

\subsubsection{Control and Management}

One of the most common applications of demand forecasts is in grid management, using a forecast to help control a storage device \cite{sossan2016atd} and optimal operational planning \cite{Lopez2019psl}. Several papers focus on PV-battery systems to help micro-grids with high penetrations of distributed energy resources \cite{mohamed2017hcd}, \cite{khan2020tee}, \cite{Zafar2018mmp}. One way to minimize the grid disturbances from high PV penetration and avoid high injection peaks is to introduce a feed-in limit, which basically caps the maximum power injection into the grid and encourages PV owners to increase their self-consumption. The feed-in limit, however, can translate into curtailment losses. To deal with this problem, \cite{Riesen2017car} present a control algorithm that aims to minimize curtailment loss and maximise self-consumption, using a linear optimization scheme that depends on the forecast data for the next 48 hour of PV production. They show that in the presence of feed-in limits, adding storage can considerably reduce curtailment losses. Litjens et al.~\cite{Litjens2018aof} also use predictive control to reduce losses due to feed-in limits . Other applications found in the literature for PV-storage systems include, aims to decrease utility bills by maximising self-consumption \cite{Johnson2018ops}, to increase PV hosting capacity \cite{hashemi2018eco}, to minimise cost in economic dispatch~\cite{bersani2006mol}, and to minimise both energy import from the local grid and energy export \cite{stephen2020ngr}.  

Other storage scenarios include trying to reduce peak load \cite{kodaira2020oes,rowe2014apr, nikolovski2018abp,bao2012bes} and controlling of electric vehicles \cite{anastasiadis2017evc}, using the vehicles as dispatchable storage units. Forecasts are used in \cite{Yunusov2018sss} for smart storage scheduling and peak reduction on LV feeders. Bennett et al. \cite{bennett2015doa} propose a scheduling system for battery energy storage (BES) with a focus on peak shaving and valley filling using 10-min data from 128 residential households. Their scheduling system comprises of generating next-day load forecasts, deriving a charge and discharge schedule based on the load forecasts, and using an online controller for making scheduler adjustments. Multiple criteria can be pursued at the same time, for example Dongol et al. \cite{Dongol2018amp} focus on peak shaving, demand smoothing and maximizing the battery utilization using model predictive control. 

As well as focusing on demand and generation, a few storage applications also consider voltage control applications. The high penetration of PV systems connected to the grid, means distribution voltage profiles are now more likely to exceed the voltage limit, potentially resulting in transformers overloading during peak production. The issue of voltage limit exceedance is thus of additional concern given the ongoing growth in solar PV systems. There has, thus, been an increased interest in problems relating to PV regulation. In this regard, \cite{Ghosh2017dvr} propose a voltage regulation technique, based on PV generation forecasts. They utilize very short-term (15 sec) PV power forecasts, using a hybrid modelling scheme based on Kalman filter theory.
Zufferey et al.~\cite{zufferey2020psf} use probabilistic short-term forecasts for constrained optimal power-flow to optimize voltage control. Hu et al.~\cite{hu2003vvc} use forecasts for Volt/Var control in distributed systems, and Wang et al.~\cite{wang2013acv} use load forecasts as inputs for a Conservation Voltage Reduction to reduce peak demand and keep voltages within regulatory standards. The authors in \cite{Kim2013ccd} propose a control method using a dynamic programming algorithm, in which the distributed generator participates in steady-stage voltage control along with switching devices. Tap changes are the traditional method for controlling the voltage, the authors of \cite{Agalgaonkar2014dvc} utilize load and irradiance forecasts to propose an optimal power coordination strategy aimed at reducing the number of tap operations, thereby minimizing the likelihood of exceeding the control limit (runaway condition) and potentially increasing the life of the tap control mechanism.

Online power management for micro-grids is presented in \cite{Mohan2016snf}. They investigate the impact of uncertainty in nodal power injections on micro-grid cost and power flow variables, using a residential feeder as a test system.

\subsubsection{Anomaly Detection}

Several papers consider how forecasts can help to identify anomalies, with the main applications being theft detection and to reduce malicious attacks. The authors in \cite{Fenza2019dma} deal with the crucial issue of theft detection in the smart grid while accommodating the concept of drift (such as a change in family size, second household etc.). The relevance of this application can be gauged from a study by the Northeast Group, LLC, which involved 125 countries, estimating that utility companies lose around USD 96 billion per annum due to nontechnical losses including fraud, theft etc. The anomaly detection strategy of \cite{Fenza2019dma} comprises of 3 steps: (1) clustering load profiles using k-means, (2) applying an LSTM model on the curves of cluster centroids and forecasting individual consumption, and (3) identifying anomalies at any given instant based on forecast errors from the previous week. Forecast errors were quantified using the RMSE, while the anomaly detection accuracy was assessed using precision and recall. Li et al. \cite{Li2019ans} develop a theft detection system for an IoT-based smart home. Their three-step methodology is based on: (1) forecasting power consumption using multiple machine learning models (MLP, RNN, LSTM, and GRU), (2) using a simple moving average for identifying the anomaly, and (3) making a final decision if the theft has occurred. Model validation was based on simulations of theft scenarios (randomly ``stealing'' energy from different time periods).

Fadlullah et al. \cite{Fadlullah2011aew} propose a probabilistic modelling methodology based on Gaussian process regression to identify malicious attack events in a smart grid, the validation of which is based on simulations. The authors state that their approach could also be used for anomaly detection, such as voltage surges and fluctuations.

In \cite{Komatsu2020pda}, forecasts are used to develop an early warning system for peak electricity consumption demand, whereby the actual weather data was linearly interpolated to match the sampling rate of the load data. However, it was not justified why linear interpolation was a suitable strategy to employ. 

The authors in \cite{mota2007lbp} use a rule-based approach and fuzzy logic concepts to predict the load behaviour after blackouts of a substation in Andorinha/Campinas/Brazil for two different blackout conditions.

\subsubsection{Flexibility Applications}

%Forecasts were also shown to be useful for flexibility applications other than storage control. 
Forecasts are useful for helping understand the effect of demand-side response by predicting the unmodified demand to compare to the actual effects after an intervention is performed as in \cite{larsen2017dre}. Similarly, \cite{priolkar2020aoc} focuses on estimating the baseline load of an LV Substation feeder in Goa (India) for implementing demand response strategies.  In contrast, Garulli et al.~\cite{garulli2015mat} forecast the actual demand-side response using the active demand (the requested change in demand).  He and Petit \cite{he2019drs} propose a scheme for demand response scheduling in a grid with high penetrations of distributed generation. 

Given a load forecast at the substation level, the approach by Ponocko et al.~\cite{ponocko2018fdf} provides a method for decomposing the forecast into the controllable and the uncontrollable components by using an ANN for disaggregation. They investigate the required percentage of users that need a smart meter, finding that a coverage of 5\% is enough to forecast the composition at the substation level with sufficient accuracy.

Pinto et al.~\cite{pinto2017mpf} use conditional kernel density estimation to generate load forecasts which feed into an optimisation that provides feasible flexibility operating trajectories that determine the storage requirements, flexible appliances or consumer preferences.

\subsection{Trading}

Peer-to-peer trading and the use of locally generated energy is expected to play a big role in future. The following papers use load forecasting for trading purposes.

Optimal solutions for multistage feeder routing problem using future loads and market prices are presented in \cite{Taghizadegan2019asc}. In  \cite{he2019mop},a real-time pricing strategy is developed. Energy trading algorithms for LV connected microgrids are discussed in \cite{feng2019hae} and a LASSO based model for household forecasts is used in \cite{kostmann2019fib} to feed blockchain designed local energy markets, which consider an auction process to match supply and demand.

\subsection{Simulating Inputs, Missing Data, Privacy protection}
\label{otherapps}
Another major application for forecasts relevant to planning, operations, and trading is to generate inputs to provide other analysis or impute unknown demand. Often forecasts are used to estimate measurements for inclusion in power flow estimation. In \cite{Korres2011SEI} an average-based forecast with Gaussian errors is used to create pseudo-observations for unmeasured loads for state estimation. Zhao et al.~\cite{zhao2020rmv} use SVR with Gaussian radial basis functions to forecast load and improve the quality of historic data for estimation of distribution system states using multi-source data, i.e., historical, online, smart meters and from Supervisory Control and Data Acquisition (SCADA) systems, as features. In \cite{Bracale2013ABB} a Bayesian forecast method is used to calculate probabilistic future steady-state analysis for a smart grid, and in \cite{Chessmore2008VPE} an ANN based model is used to estimate the voltage profile via a power flow program. Finally, \cite{hermanns2020eod} use a simple average load forecast of the past weeks with a correction of the forecast values as new load values become available. The forecast is used for a grid state forecast software tool.

Forecasts are also used to fill in or estimate missing data, as in Borges et al.~\cite{borges2020etm} who use short-term forecasting models with adjusted features (using future values) to impute missing data for primary substations. They find that, while it differs for different data sets, generally the model with access to historical data, meteorological data and also data from other neighboring substations can improve over using only subsets of these features. Zhou et al.~\cite{zhou2020blb} use LSTM for load forecasts in their approach to providing harmonic state estimation using regression analysis for power flow calculations and sparse Bayesian learning. Finally, in \cite{wang2020rnl}, the authors consider decomposing the demand of LV substations into traditional load, flexible load and distributed generation components. An exponential smoothing load forecast is used as an estimate for the load component of the decomposition. 

Huyghues-Beaufond et al. investigate the effects of data cleansing on the forecast accuracy of LV/MV feeders of UK Power Networks \cite{huyghues-beaufond2020raa}. Their methodology consists of the following steps. First, outliers are detected using an automatic procedure which combines the Tukey labelling rule \cite{Tukey} and the binary segmentation algorithm. Next, various approaches for missing value imputation are investigated, including unconditional mean, hot-deck via \textit{k}-nearest neighbour and Kalman smoothing. Feed-forward deep neural networks for day-ahead forecast at hourly resolution are developed to assess the performance of the cleansing method. 
The proposed data cleansing framework efficiently removes outliers from the data, and the accuracy of forecasts is improved. It is found that hot deck (\textit{k}-NN) imputation performs best in balancing the bias-variance trade-off for short-term forecasting.

Chen et al. \cite{chen2014ita}  use Chebyshev’s inequality to identify inaccurate observations and use feature curves to restore these data points. It is shown that correction of data improves the overall forecast accuracy. 

Privacy protection is important for different applications, and in \cite{boustani2017sgp} the methods are developed to protect household privacy whilst preserving load profile correlations between forecasts and actuals. 
Similarly, in  \cite{hou2020anp} privacy-preservation via a model randomization scheme 
consisting of a forecasting phase and a reporting phase is presented. The approach is currently limited to multivariate polynomials. However, many different statistical and machine learning methods such as ARIMA, MLR, RBF and ANN can be represented using modifications.
Finally,  a security and privacy preserving scheme by limiting the connectivity of home area networks with the electric grid is presented in \cite{abdallah2017lsa}.

\subsection{Summary}
\label{sec:summary}

Although forecasts are essential for the presented applications, with the exception of the storage control papers there is often very little focus on the methods or the forecast errors within the reviewed papers. In many cases, the forecast models used are not presented, and in others, it is not clear if a forecast has been generated, or simply the actual future values are used as proxies for real forecasts. When actual forecasts are developed for the applications, most use naive methods such as basic averages, or similar day methods. When forecast errors are used, they usually rely on basic Gaussian assumptions.

Given the volatility of distribution level demand, it is surprising that relatively few of the application papers utilise probabilistic forecasts, despite some cases such as \cite{nikolovski2018abp} (using an adaptive neuro-fuzzy inference system), \cite{zufferey2020psf} (probabilistic kNN), and \cite{pinto2017mpf} (conditional KDE). Lilla et al. \cite{Lilla2020dsl} consider day-ahead scheduling of a battery energy storage system with PV and, as stated by the authors, the uncertainty in forecasts was ignored in the analyses, and the prices of exchanges with the utility grid are assumed to be predefined.

An important gap apparent from this review is that there are very few examples of the impact and role the forecast's accuracy has on the outputs of the application. Most papers do not include a benchmark or different models with which to compare how the accuracy of the model affects the performance of the application. Without these investigations and comparisons, it is difficult to assess the value or importance of the forecast's quality. A few examples, including \cite{stephen2020ngr}, \cite{alasali2020acs} have shown that improved forecast accuracy can improve the performance of the application. Further \cite{kodaira2020oes} shows that using prediction intervals improves peak load reduction compared to some basic point forecast methods. However, there are no detailed studies of performance changes of probabilistic versus point forecasts. Understanding whether probabilistic forecasts offer significant improvements over point forecasts can help make decisions of whether probabilistic methods are worth the increased computational costs.

\section{Discussion}
\label{sec_discussion}

The area of LV load forecasting has garnered growing attention in the last few years, which is evident from a significant increase in the number of publications, datasets, methodologies and applications, as discussed in this paper. While these recent developments are helping to move towards increasingly efficient, digital, and easy to monitor local energy systems, we are also faced with new challenges and subsequent opportunities. In this section, we discuss some of these challenges and share our views and recommendations. We also discuss gaps in the literature and possible (and desirable) future directions. 

\subsection{Recommendations}

\begin{itemize}
	\item \textit{Tackling single-source data bias} - A vast majority of the literature on modelling smart electricity meter time series have employed the Irish CER dataset, potentially because it was one of the first open-source repositories of such a dataset. \textbf{Recommendations:} to better gauge the practical scalability of models and the generalizability of findings reported using the Irish CER data, external validation using other datasets is needed. An interesting line of study would be to train models using say, the Irish CER data, but then validate and compare the models using multiple datasets from different sources (a validation approach that is sometimes used in the area of medical diagnostics where data from multiple independent cohorts are used).  Here, a large suitable, more recent data set is the UK Low Carbon London trial~\cite{UK2014ulc} that has been archived for better analysis and referenced in the Monash Time Series Forecasting Archive~\cite{godahewa2021mts} in a pre-processed and raw version.
	
	\item \textit{Towards clarity in problem definition} -  To improve readability, it will help if the papers clearly and concisely describe the problem statement upfront, before delving into the intricacies of the methodology. From some papers, even a plot of the time series, or details regarding the forecast horizon, were missing. \textbf{Recommendations:} information such as forecast horizon, level of aggregation (one household, feeder, region, etc.), a plot of the time series, sampling rate, error measures, data source, pre-processing steps (for handling missing observations and anomalous load profiles due to public holidays etc), and forecasting scheme (direct vs iterative) must be provided. 
	
	\item \textit{Need for benchmarks and robust validation}: Numerous articles did not employ benchmarks to compare their models against (they typically compared different variants of their presented approach). For HV load modelling, a range of naive benchmarks (such as seasonal random walk, seasonal moving average) and sophisticated benchmarks (such as SARMA, HWT exponential smoothing, ANNs) are typically used. However, this is not the case for papers dealing with LV load modelling. \textbf{Recommendations:} a comparison with an existing methodology proposed in the literature for a similar problem/data is needed, or at least some of the naive and sophisticated benchmarks specified above could be employed such as those used in \cite{Haben2019stl}. Also, how the data is split into train and test sets should be clearly stated. Very often hyper-parameter tuning was missing or hyper-parameters were chosen based on the test set, whereas a validation set should have been defined and be separate from the test set. The test dataset should be sufficiently large and representative.
	
	\item \textit{Modelling uncertainty due to weather}: A significant number of the papers that we surveyed used no weather information. A few papers that used weather variables, used weather actuals, thereby under-reporting the forecast errors. Only a handful of papers used weather predictions, but typically for one weather variable that was obtained from one weather station. \textbf{Recommendations:} weather ensemble predictions (ideally for different weather variables obtained from multiple weather stations) need to be used while modelling the LV load. Weather ensembles from weather stations are prone to be biased and under-dispersed.  There have been significant advancements in the area of numerical weather predictions in the last decade, unfortunately, these advancements have not yet translated into improved LV load forecast accuracy.  
	
	\item \textit{Moving towards probabilistic forecasting} - Research on LV load forecasting has tended to focus on generating point forecasts. However, the time series at the LV level exhibit considerable variability along with seasonality. \textbf{Recommendations:} (1) The aim of modelling should be to generate and evaluate probabilistic forecasts (and not just focus on a point estimate or a pre-specified quantile). (2) Studies focusing on peak LV load forecasts should use modified error measures that avoid the double penalty. An interesting line of work would be on probabilistic peak forecast evaluation at the LV level. (3) Model estimation based on in-sample probabilistic forecast error measures needs to be considered. Crucially, studies should report the model hyper-parameters along with details of the estimation framework, to help improve reproducibility. (4) A plot of probabilistic forecast accuracy (ideally quantified using a strictly proper scoring rule) versus forecast horizons should be provided, as different models may perform well at different horizons. (5) If needed, statistical significance tests should be used for model comparison.
	
	\item \textit{Improving access to literature}: despite generally a trend towards more open research by using preprint servers and Open Access journals, currently, the majority of journals in this area are not Open Access. This makes it expensive especially for industry, as unlike universities, they typically don't have subscriptions with the publishers. Further, several papers were found only in non-English speaking languages limiting the contribution to the international forecasting community. \textbf{Recommendations:} more journals should allow more explicitly the sharing of preprints for more open research. Research funding agencies should fund and even encourage open access publication (public money for public research), despite being often a considerably more expensive option. National funding agencies should further encourage the exchange with international research communities by encouraging publications, in both native languages and English.
	
	\item \textit{Becoming a supportive community}: a considerable proportion of papers used closed data sets \emph{and} closed code, while in many other communities the code is typically shared along with the paper. However, the papers reviewed in this work typically did not publish the modelling and algorithmic code. \textbf{Recommendations:} sharing the code or even the pseudo-code (along with model hyper-parameters) will increase the likelihood of adoption of the proposed methodologies. 
	
	\item \textit{Applications}: for many applications, naive forecast methods are used or an accurate forecast is assumed without much investigation into what role the forecast accuracy plays in the application. Often a forecasting method is not benchmarked at all, or worse still, no forecast is given. These scenarios mean that there is no clarity of the effect of the forecast. If a forecast has minimal impact then practitioners may apply far too much time and effort to develop a very accurate model. In contrast, if an accurate forecast is key, the application may be abandoned if the required results are not delivered. \textbf{Recommendations:} The accuracy of the forecast model and the comparison to a benchmark must be presented. This will allow a proper assessment of the forecast accuracy and the influence of forecast errors on the performance within the application. 
	
	\item \textit{Privacy}: one of the main obstacles to obtaining shared datasets from LV networks and households is data privacy, i.e. personal and life-style information can be deduced from the energy usage, and operational details for LV networks usually cannot be exposed due to safety or commercial reasons. Privacy-protecting analytics that mitigates those issues, such as differential privacy, federated learning and other methods are already in development and used (e.g. in life-sciences datasets to protect privacy) but not much is applied to energy data (we listed a few examples of papers that considered privacy protection measures in \ref{otherapps}). \textbf{Recommendations:} developing and evaluating methods for privacy protection in LV datasets will eventually enable opening and sharing (or at least modelling) of many more datasets. 
	
	\item \textit{Use of Computational Intelligence Models}: there is a recent trend to use computationally complex models from the deep learning domain on all kinds of problems with abundant data. With a plethora of papers on novel deep learning architectures appearing as preprints and the major machine learning conferences, it is possible to pick novel algorithms and mechanisms, apply them to the LV load forecasting problem and with sufficient tuning achieve good performance on a test set. Given the above challenges (dataset bias and missing benchmarks), it is therefore hard to judge the contribution of a novel approach. Compared to statistical models, current computational models are generally less interpretable and computationally expensive, making them less attractive to industrial and commercial organisations, as they are more costly to run, less ecologically sustainable, and harder to "trust". \textbf{Recommendations:} Instead of comparing computational approaches only to other computational approaches or to simple benchmarks, novel approaches should be compared to strong statistical models like SARMA and HWT exponential smoothing. Model evaluation should be done following cross-evaluation and a large enough test set (separate from that used for model tuning) should be used for comparison to other strong benchmarks to ensure generalisation. The evaluation should further consider not only forecasting errors, but also computational complexity by reporting e.g. running times or even energy consumption and qualitative properties like model interpretability.

\end{itemize}

\subsection{Bridging the gap and future directions}
From this wide coverage of current work on methods and applications, we have chosen the most pressing and important open problems and identified gaps and directions for future research.

Firstly, it should be noted that no individual method can currently be considered state-of-the-art in LV load forecasting, i.e. no method has been shown to consistently provide significantly improved results (relative to either application or appropriate metrics) over any other methods. In the literature, there are only a few wide-ranging comparisons, and even these typically only focus on a few models within a specific family of methods (neural networks or regression models, for example). As discussed in Section \ref{sec:LVLF-applications}, understanding which forecasting methods may be most appropriate or optimal for each application is even more unclear. Minimal focus is applied on the forecasts, which are simply treated as required inputs rather than  essential components of the respective problems. In the early development of the technical solutions to these applications, this may be understandable, however as the area matures, focus on the forecasts will be essential and will require detailed analysis of their role and structure. 

A major motivation for LV level forecasts is that in a smart grid, LV networks open up a wide range of
solutions and opportunities for new applications as shown in Section \ref{sec:LVLF-applications}. Not only
does this mean that algorithms must be tested with regard to how viable they are for the
particular application, but they might also require new and novel inputs, not usually required at
HV level, as demonstrated in Section \ref{subsec:novel_var}. Responses to explanatory variables are also largely unexplored at the LV level, as shown in Section \ref{subsec:expl_var}. 
Furthermore, given the huge recent progress in numerical weather prediction, an interesting avenue of future research would be
\begin{itemize}
	\item to use post-processed weather ensemble predictions to generate multi-step probabilistic forecasts of load at different levels of the LV hierarchy. 
	\item to investigate how different forecasting skills for renewable generation and skills for demand vary over different time-horizons and spatial resolutions combined.
\end{itemize}

Moreover, given that LV networks' loads are much more volatile, the development of probabilistic forecasts is an obvious direction for future work, together with developing error metrics suitable for training and use of those models, which are able to cope with the double penalty effect discussed in Section \ref{subsec:eval}. 

Another major topic, given the lack of data and benchmarks, is collaboration and sharing of datasets. There is an urgent need for the development of privacy-protecting analytics that could overcome commercial, safety and privacy concerns -  major barriers toward opening more LV load datasets.
Several open problems are related to this area: 
\begin{itemize}
	\item robust creation of synthetic, `look alike' data,  (e.g. by adding noise to datasets,  applying differential privacy, etc.);
	\item development of new or adaptation of existing  privacy protection techniques, such as federated learning etc.;
	\item quantification of how increasing privacy protection influences the accuracy of forecasting methods;
	\item exploration of adversarial techniques to keep those mitigation methods resilient ( i.e. showing that the data cannot be used in combination with other datasets to deduce identity etc.).
\end{itemize}
Finally, there is a need for developing pragmatic methods to achieve explainability of AI and ML methods in the LV load forecasting context. This will require  working interdisciplinarily with end users (power and system engineers), and developing a deeper understanding of the operational environment by AI scientists.

\section*{Acknowledgment}
Marcus Voss has been partially funded by the German government under funding ref. number 03SIN539 (WindNODE).

\bibliographystyle{plain}
\bibliography{references,georgios,datasetrefs}
\end{document}